\documentclass[journal]{IEEEtran}
\usepackage{amsmath,amssymb,amsthm,lipsum} 
\usepackage{cite}
\usepackage{algorithmic}
\usepackage{algorithm}
\usepackage{graphicx}
\usepackage{textcomp}
\usepackage{stfloats}
\usepackage{booktabs}
\usepackage{bm}
\usepackage{subfigure}
\usepackage{setspace}
\usepackage{color}
\usepackage[T1]{fontenc}

\ifCLASSINFOpdf
\else
\fi
\theoremstyle{Theorem}
\newtheorem{theo}{Theorem}
\newtheorem{theoremmaindescription1}[theo]{Theorem}
\newtheorem{theoremmaindescription2}[theo]{Theorem}

\theoremstyle{remark}
\newtheorem{rmk}{Remark}
\newtheorem{rmk2}[rmk]{Remark}
\newtheorem{rmk3}[rmk]{Remark}

\newtheorem{rmk5}[rmk]{Remark}

\theoremstyle{Definition}
\newtheorem{def1}{Definition}
\newtheorem{def2}[def1]{Definition}
\newtheorem{def3}[def1]{Definition}

\theoremstyle{Lemma}
\newtheorem{lemma1}{Lemma}

\newtheorem{lemma3}[lemma1]{Lemma}
\newtheorem{lemma4}[lemma1]{Lemma}
\newtheorem{lemma5}[lemma1]{Lemma}
\newtheorem{lemma6}[lemma1]{Lemma}
\newtheorem{lemma7}[lemma1]{Lemma}

\theoremstyle{Corollary}
\newtheorem{Corollary1}{Corollary}
\newtheorem{Corollary2}[Corollary1]{Corollary}
\newtheorem{Corollary3}[Corollary1]{Corollary}
\newtheorem{Corollary4}[Corollary1]{Corollary}
\newtheorem{Corollary5}[Corollary1]{Corollary}
\newtheorem{Corollary6}[Corollary1]{Corollary}
\newtheorem{Corollary7}[Corollary1]{Corollary}

\begin{document}
%
\title{ Compressive Spectrum Sensing\\ Using Sampling-Controlled \\
Block Orthogonal Matching Pursuit}
%
%
%

\author{Liyang~Lu,~\IEEEmembership{Graduate~Student~Member,~IEEE,}
        Wenbo~Xu,~\IEEEmembership{Member,~IEEE,}\\
        Yue~Wang,~\IEEEmembership{Senior Member,~IEEE,}
        and Zhi~Tian,~\IEEEmembership{Fellow,~IEEE}
\thanks{
	This work was supported in part by the National Natural Science Foundation of China under Grant 61871050, the US National Science Foundation under Grants 2136202 and 2128596, and the Virginia Research Investment Fund CCI under Grant 223996.
	
	L. Lu and W. Xu are with the Key Lab of Universal
Wireless Communications, Ministry of Education, Beijing University of
Posts and Telecommunications.

Y. Wang and Z. Tian are with the Department of Electrical and Computer
Engineering, George Mason University, Fairfax, VA.

W. Xu (e-mail: xuwb@bupt.edu.cn) and Y. Wang (e-mail:
ywang56@gmu.edu) are the corresponding authors.}
}

%
%

\markboth{IEEE TRANSACTIONS ON COMMUNICATIONS}%
{Shell \MakeLowercase{\textit{et al.}}: Bare Demo of IEEEtran.cls for IEEE Journals}
%



\maketitle

\begin{abstract}
This paper proposes two novel schemes of wideband compressive spectrum sensing (CSS) via block orthogonal matching pursuit (BOMP) algorithm, for achieving high sensing accuracy in real time. These schemes aim to reliably recover the spectrum by adaptively adjusting the number of required measurements without inducing unnecessary sampling redundancy. To this end, the minimum number of required measurements for successful recovery is first derived in terms of its probabilistic lower bound. Then, a CSS scheme is proposed by tightening the derived lower bound, where the key is the design of a nonlinear exponential indicator through a general-purpose sampling-controlled algorithm (SCA). In particular, a sampling-controlled BOMP (SC-BOMP) is developed through a holistic integration of the existing BOMP and the proposed SCA. For fast implementation, a modified version of SC-BOMP is further developed by exploring the block orthogonality in the form of sub-coherence of measurement matrices, which allows more compressive sampling in terms of smaller lower bound of the number of measurements.
Such a fast SC-BOMP scheme achieves a desired tradeoff between the complexity and the performance. Simulations demonstrate that the two SC-BOMP schemes outperform the other benchmark algorithms.
\end{abstract}


%
\IEEEpeerreviewmaketitle

\section{Introduction}
\label{introduction}
%
%
%
%

\IEEEPARstart{I}{n} cognitive radio network (CRN) \cite{33,34,35,51}, spectrum resources become more and more scarce, leading to the emergence of spectrum sensing \cite{36,37} in dynamically providing the spectrum occupancy for secondary users (SUs). The spectrum holes can thus be used by SUs without causing any interference to primary users (PUs). It is known that wider spectrum can enable CRN to obtain more opportunities of access \cite{16,54}, inspiring the interests of many scholars in wideband spectrum sensing.

The main challenge of wideband spectrum sensing is the high sampling rate, which results in huge resource consumption. Fortunately, the compressive sensing (CS) technology with sub-Nyquist-rate sampling has been used to address this challenge, which is called compressive spectrum sensing (CSS) \cite{39,21,38,52,53,55}. Among various CSS algorithms, block orthogonal matching pursuit (BOMP) is a typical one that effectively recovers the spectrum with block sparsity in an iterative manner \cite{17}. There are two major advantages highlighting the wide use of BOMP in practical applications: firstly, BOMP achieves much better recovery performance than some conventional CSS algorithms, such as orthogonal matching pursuit (OMP) and orthogonal least squares (OLS), in reconstructing underlying signals with block structure; secondly, as a greedy algorithm, BOMP has lower computational complexity than convex optimization algorithms or some other greedy algorithms, e.g., block OLS.

\subsection{Related Works}

	However, the existing BOMP still causes nontrivial complexity as the problem size goes large in wideband scenarios. It is thus necessary to further speed up spectrum sensing \cite{27,23,24,28} for fast implementation of wideband sensing which leaves more computation resources and time duration for the follow-up data transmission on the detected spectrum opportunities.

	Recent studies \cite{45,46,47} propose adaptive and efficient spectrum sensing schemes, but they ignore block structure of the sparse spectrum \cite{17}. It has been proved that the use of block sparsity results in faster and more accurate sparse recovery \cite{4,49,2}.  
	Moreover, BOMP enjoys the merits of fast and accurate sensing performance \cite{2,4}, which are the merits that CSS calls for. However, to the best of our knowledge, it is still lack on both the theoretical analysis and practical algorithm design for BOMP in CSS. Therefore, to fill the aforementioned gap, this work is motivated to investigate the fundamental limits and algorithmic designs on the BOMP-based CSS methods.

	Note that the CSS using BOMP can be accelerated by appropriately reducing the number of measurements, since the computational complexity of BOMP algorithm is mainly determined by the number of measurements \cite{19,20,2,8,14}. 
In \cite{20}, the authors derive the bound on the necessary number of measurements for the greedy algorithm, which indicates that if $M\geq4kd\ln(2N/\omega_1)$, the probability of exact sparse recovery using the greedy algorithm is no lower than a given threshold in the noiseless case, where $\omega_1$ is a constant. 
The recent study \cite{14} improves the bound in \cite{20} to $M\geq2kd\ln(N/\omega_2)$ for the greedy algorithm, where $\omega_2$ is also a constant.

Although the theoretical lower bound for the number of measurements is continuously improved, a gap still exists between the empirical necessary number of measurements and the theoretical results. That is, the empirical number for reliable recovery is much smaller than the theoretical bounds \cite{14}. 
	When humans have to use the theoretical bounds as the guideline to determine the number of measurements, such gap usually leads to a wastage on the excessive measurements collected in practical CSS. This is because these theoretical bounds are too loose for CSS application. Therefore, a key task is to shrink this gap. 

Moreover, the noise effect is usually ignored in existing works \cite{20,14} for simplicity of analysis. Considering the fact that noise always exists in practical scenarios, it is necessary to develop a tight lower bound of the number of measurements by taking consideration of the noise effect for CSS.	

\subsection{Our Contributions}

To reduce the computational cost of BOMP in CSS scenarios without sacrificing its recovery accuracy, this paper proposes two novel schemes called general sampling-controlled BOMP (SC-BOMP) and fast SC-BOMP respectively. In these schemes, two sampling-controlled algorithms (SCAs) are developed to dynamically adjust the number of measurements to an appropriate level, where tighter lower bounds on the number of minimum measurements in the noisy scenarios are derived which in turn shed lights on the design of efficient iterative algorithms. The proposed SCAs promote the practical usage of the theoretical analyses on the necessary number of measurements, and the combinations of the SCAs and BOMP, i.e., the SC-BOMP schemes, perform well in the actual CSS application.
The contributions of this paper are summarized as follows.

\begin{enumerate}
	\item This work derives the lower bounds of the necessary number of measurements of BOMP, which ensure the probability of reliable recovery exceeding a given constant in both noiseless and noisy cases. By using the $\ell_{2,\infty}$-norm formulated coherence in a block manner, our developed results are tighter than those in \cite{20} and \cite{14} based on our extended support selection condition.
	Furthermore, to the best of our knowledge, this is the first work developed to reveal the effect of noise power on the necessary number of measurements for the BOMP algorithm.
	
	\item Two sampling-controlled algorithms (SCAs), i.e., the general SCA and the fast SCA, are proposed based on the derived bounds. Specifically, in the general SCA, the bounds on the necessary number of measurements are further reduced to a tight level by using the widely adopted exponential function indicator. The tightness means that BOMP may not perform 100\% spectrum detection even if the number of measurements decreases slightly, and the gap between the theoretical bound and the empirical result has been greatly reduced. In the fast SCA, this gap is further reduced by utilizing the block orthogonality of the measurement matrix. Note that the  block orthogonality means that the sub-coherence of the measurement matrix is equal to 0, leading to the improvements of the bounds on the necessary number of measurements. The fast SCA  can still enable BOMP to realize 100\% sensing, but it requires higher signal-to-noise ratio (SNR) than that of the general SCA.
	
	\item Two CSS schemes, i.e., the general SC-BOMP and the fast SC-BOMP, are proposed to realize reliable and real-time sensing. It can be inferred that the general SC-BOMP is a combination of the general SCA and BOMP, while the fast SC-BOMP is based on the fast SCA. They effectively utilize the theoretical bounds on the number of measurements for practical CSS applications, and achieve the desired sensing performance at affordable computing resources and running time. That is, in low SNR environments, the schemes perform reliable CSS, while in high SNR environments, the schemes obtain 100\% probabilities of detection without the waste of computing resources. 
	
	\item The numerical simulations verify our theoretical results that our derived bound on the necessary number of the measurements is lower than the existing ones. The CSS simulations indicate that our proposed SC-BOMP schemes achieve high sensing accuracy by adaptively selecting the number of required measurements, resulting in reliable performance and effective saving of computing resources. Meanwhile, the fast SC-BOMP scheme utilizing block orthogonality of the measurement matrix performs faster spectrum sensing than the general SC-BOMP scheme at the cost of only a slight accuracy loss.
\end{enumerate}

The rest of this paper is organized as follows. Section II introduces notations, system model, some definitions and useful theoretical warm-ups, which facilitates the subsequent study of the lower bound for necessary number of measurements in Section ${\rm \uppercase\expandafter{\romannumeral3}}$. In Section ${\rm \uppercase\expandafter{\romannumeral4}}$, the two versions of SC-BOMP schemes are proposed. Simulation results are presented in Section ${\rm \uppercase\expandafter{\romannumeral5}}$, followed by conclusions in Section ${\rm \uppercase\expandafter{\romannumeral6}}$.

\section{Preliminaries}
\subsection{Notations}
Denote vectors by boldface lowercase letters, e.g., $\mathbf{r}$, and matrices by boldface uppercase letters, e.g., $\mathbf{D}$. The $i$-th element of $\mathbf{r}$ is denoted as $\mathbf{r}_i$ and $\mathbf{r}^t$ represents $\mathbf{r}$ in the $t$-th iteration. The element in the $i$-th row and $j$-th column of matrix $\mathbf{D}$ is denoted as $\mathbf{D}_{ij}$, and $\mathbf{D}_i$ is the $i$-th column of $\mathbf{D}$. $\mathbf{D}^T$ represents the transpose of matrix $\mathbf{D}$. Letting $k$ be the block sparsity, $\mathbf{\Omega}\in \mathcal{R}^{k}$ denotes the set containing the indices of nonzero blocks of a sparse signal. The sub matrix $\mathbf{D}_{\mathbf{\Omega}}\in R^{M\times kd}$ consists of matrix blocks corresponding to $\mathbf{\Omega}$ and the same as the sub vector $\mathbf{x}_{\mathbf{\Omega}}$, where $M$ is the number of measurements. If $\mathbf{\Omega}$ is replaced by another set $\mathbf{S}$, then the same is true as before. $\bar{\mathbf{\Omega}}$ is the complementary set of $\mathbf{\Omega}$ and $\mathbf{\Omega}\setminus \mathbf{S}=\{i|i\in\mathbf{\Omega},i\notin \mathbf{S}\}$. If the objective in $|\cdot|$ is a numerical value, $|\cdot|$ means the absolute value of its target and if the objective is a finite set, $|\cdot|$ denotes its cardinality. $||\cdot~||_{0}$, $||\cdot~||_{1}$, $||\cdot~||_{2}$ and $||\cdot~||_{\infty}$ represent the $\ell_0$, $\ell_1$, $\ell_2$ and $\ell_{\infty}$ norms of their targets respectively. $||\cdot~||_{2,0}$, $||\cdot~||_{2,1}$ and $||\cdot~||_{2,\infty}$ are the $\ell_{2,0}$, $\ell_{2,1}$ and $\ell_{2,\infty}$ mixed norms of their targets. The spectral norm of $\mathbf{D}$ is represented by $||\mathbf{D}||_2=\sqrt{\lambda_{\max}(\mathbf{D}^T\mathbf{D})}$, where $\lambda_{\max}(\mathbf{D}^T\mathbf{D})$ is the largest eigenvalue of $\mathbf{D}^T\mathbf{D}$. For the index set $\mathbf{S}^t$ in the $t$-th iteration, if $\mathbf{D}_{\mathbf{S}^t}$ has full column rank, then $\mathbf{P}_{\mathbf{S}^t}=\mathbf{D}_{\mathbf{S}^t}(\mathbf{D}_{\mathbf{S}^t}^T\mathbf{D}_{\mathbf{S}^t})^{-1}\mathbf{D}_{\mathbf{S}^t}^T$ denotes the projection onto the span of $\mathbf{D}_{\mathbf{S}^t}$. $\mathbf{P}^{\bot}_{\mathbf{S}^t}=\mathbf{I}-\mathbf{P}_{\mathbf{S}^t}$ is the projection complement of the span of $\mathbf{D}_{\mathbf{S}^t}$, where $\mathbf{I}$ is the identity matrix. $\mathbb{E}(\cdot)$ and ${\rm P}(\cdot)$ denote the expectation and probability of their targets respectively. $\mathbf{E}(\cdot)$ represents a random event. $\lceil\cdot\rceil$ denotes the ceiling operation.

\subsection{System Model}

\begin{figure}
	\centering
	\includegraphics[scale=0.4]{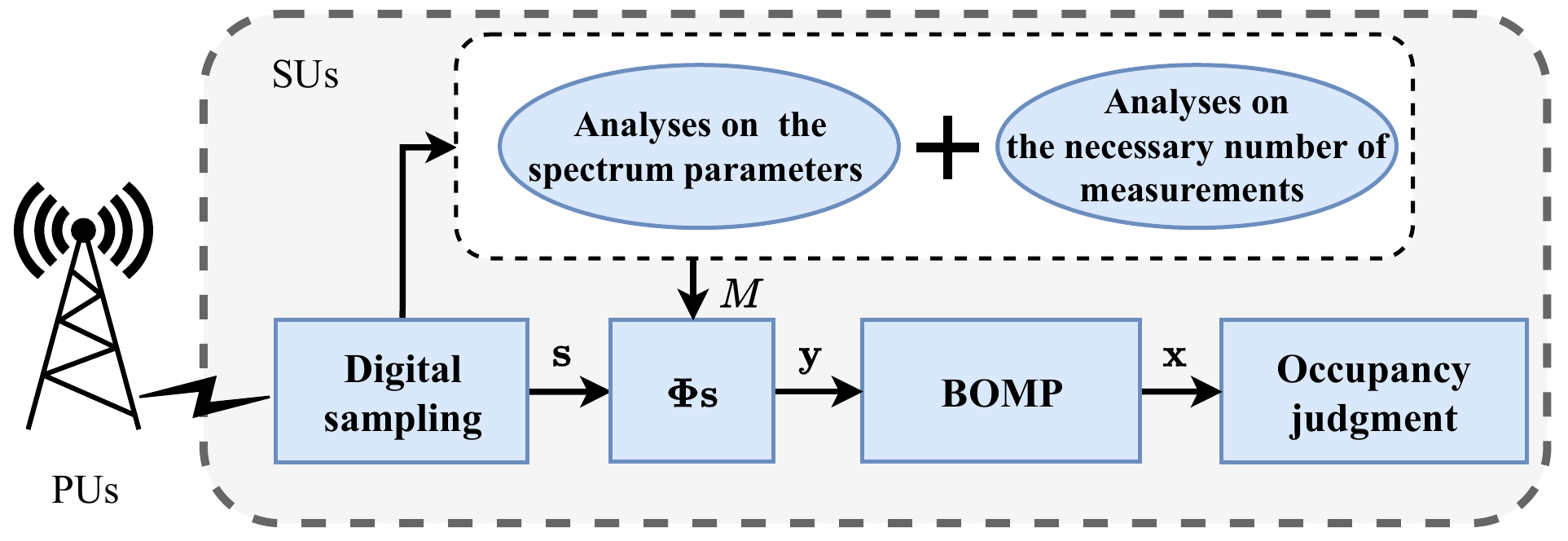}\\
	\caption{System model of the proposed CSS schemes.}\label{framework2}
\end{figure}

 As shown in Fig. \ref{framework2}, a wideband spectrum signal that SUs receive from PUs is $\mathbf{s}\in \mathcal{R}^{N}$, which is sparse based on a certain basis $\mathbf{\Psi}\in \mathcal{R}^{N\times N}$ \cite{39}, i.e., $\mathbf{s}=\mathbf{\Psi}\mathbf{x}\in \mathcal{R}^{N}$ with $||\mathbf{s}||_0\ll N$. Meanwhile, $\mathbf{s}$ can be undersampled by a compression matrix $\mathbf{\mathbf{\Phi}}\in \mathcal{R}^{M\times N}$ with $M\ll N$ . Due to the additive noise in practical CSS scenario, the system model is formulated by \cite{16}
 \begin{equation}\label{SYSTEMMODEL}
 	\mathbf{y}=\mathbf{\Phi} \mathbf{s}+\mathbf{n}=\mathbf{\Phi}\mathbf{\Psi}\mathbf{x}+\mathbf{n}=\mathbf{D}\mathbf{x}+\mathbf{n},
 \end{equation}
 where $\mathbf{y}\in \mathcal{R}^{M}$ is the measurement vector, $\mathbf{D}=\mathbf{\Phi}\mathbf{\Psi}\in \mathcal{R}^{M\times N}$ is the effective measurement matrix and $\mathbf{n}\in \mathcal{R}^{M}$ represents the additive noise satisfying $||\mathbf{n}||_2\leq\epsilon$. It is proved that the restricted isometry and mutual incoherence properties of a given Gaussian matrix are satisfactory for reliable sparse recovery \cite{44}. As a result, Gaussian measurement matrix is widely used in CSS \cite{16,21}.
 
 Different from conventional CSS \cite{39,16,22} that the number of measurements $M$ is fixed, this work can set a sufficient but not superfluous number according to the theoretical results, and thus save the transmission burden and recovery cost. Accordingly, the overall CSS system enjoys the reduced cost of computing resources and real-time implementation advantages.
 
 Further, the spectrum in CRN usually appears with block sparsity \cite{17}. The block sparse spectrum $\mathbf{x}$ is defined as
\begin{equation}\label{block sparsemodel}
\mathbf{x}=[\underbrace{\mathbf{x}_1\cdots  \mathbf{x}_d }_{\mathbf{x}^T[1]} \underbrace{\mathbf{x}_{d+1}\cdots \mathbf{x}_{2d}}_{\mathbf{x}^T[2]}\cdots \underbrace{\mathbf{x}_{N-d+1}\cdots \mathbf{x}_{N}}_{\mathbf{x}^T[N_b]}]^T,
\end{equation}
where $d$ is block length, $N=N_bd$ and $\mathbf{x}[i]\in \mathcal{R}^{d}$  $(i\in\{1,2,\cdots,N_b\})$ is the $i$-th block of $\mathbf{x}$.
Denoting
\begin{equation}\label{l20}
||\mathbf{x}||_{2,0}=\sum^{N_b}_{i=1}\mathcal{I}(||\mathbf{x}[i]||_2>0)
\end{equation}
with the binary indicator function $\mathcal{I}(\cdot)$. Then,
if $\mathbf{x}$ is $k$ block sparse, $||\mathbf{x}||_{2,0}\leq k$. Accordingly, the measurement matrix is rewritten as a concatenation of $N_b$ column blocks, i.e.,
\begin{equation}\label{matrixblock}
\mathbf{D}=[\underbrace{\mathbf{D}_1\cdots  \mathbf{D}_d }_{\mathbf{D}[1]} \underbrace{\mathbf{D}_{d+1}\cdots \mathbf{D}_{2d}}_{\mathbf{D}[2]}\cdots \underbrace{\mathbf{D}_{N-d+1}\cdots \mathbf{D}_{N}}_{\mathbf{D}[N_b]}],
\end{equation}
where $\mathbf{D}[i]\in \mathcal{R}^{M\times d}$ is the $i$-th block of $\mathbf{D}$ $(i\in\{1,2,\cdots,N_b\})$. Assume that the measurement matrix $\mathbf{D}$ is an $M\times N$ matrix with $M=Ld$ and $N=Rd$, where $d$ represents the block length, i.e., the number of measurements and the number of atoms are integral multiples of the block length $d$.
In this paper, the BOMP algorithm \cite{4} given in Algorithm \ref{alg:22} is utilized to recover the sparse spectrum $\mathbf{x}$ from the measurement vector $\mathbf{y}$.
	Finally, based on the reconstructed $\mathbf{x}$, SUs can identify the idle spectrum.

\begin{algorithm}
	\renewcommand{\algorithmicrequire}{\textbf{Input:}}
	\renewcommand{\algorithmicensure}{\textbf{Output:}}
	\caption{Block orthogonal matching pursuit}
	\label{alg:22}
	\begin{algorithmic}[1]
		\REQUIRE $\mathbf{D}, \mathbf{y}$, block sparsity level $k$ and block length $d$
		\ENSURE $\hat{\mathbf{S}}\subseteq \mathbf{\Omega}$ and $\hat{\mathbf{x}}\in\mathcal{R}^{N}$
        \STATE $\mathbf{Initialization:}$ $t=0$, $\mathbf{r}^0=\mathbf{y}$, $\mathbf{S}^0=\emptyset$, $\mathbf{x}^0=\mathbf{0}$
        \WHILE {$t< k$ or not converged}
		\STATE Set $i^{t+1}=\mathop{\arg\max}\limits_{j\in\{1,\cdots,N_b\}\backslash\mathbf{\Omega}^{t}}||\mathbf{D}^T[j]\mathbf{r}^{t}||_2$
		\STATE Augment $\mathbf{S}^{t+1}=\mathbf{S}^{t}\cup{\{i^{t+1}\}}$
		\STATE Estimate $\mathbf{x}^{t+1}[j]=\mathop{\arg\min}\|\mathbf{y}-\sum\limits_{j\in\mathbf{S}^{t+1}}\mathbf{D}[j]\mathbf{x}^{t+1}[j]\|_2$
        \STATE Update $\mathbf{r}^{t+1}=\mathbf{y}-\mathbf{D}\mathbf{x}^{t+1}$
        \STATE $t=t+1$
        \ENDWHILE
		\STATE \textbf{return} $\hat{\mathbf{S}}=\mathbf{S}^{t}$ and $\hat{\mathbf{x}}=\mathbf{x}^{t}$
	\end{algorithmic}
\label{bomp}
\end{algorithm}

\subsection{Useful Definitions, Lemmas and Corollaries}
In order to facilitate the following analysis, some definitions, lemmas and corollaries are given first.
\begin{def1} The metric that measures the disparity of a general signal $\mathbf{x}$ is given by $||\mathbf{x}_\mathbf{S}||_{2,1}^2\leq ||\mathbf{x}_\mathbf{S}||_2^2|\mathbf{S}|, \quad\forall \mathbf{S}\subseteq\mathbf{\Omega}$ \cite{1,2}.
\label{def1}
\end{def1}

\begin{def2}
The matrix coherence $\mathbf{D}$ is defined as $\mu=\max_{i,j\neq i}|\mathbf{D}_i^T\mathbf{D}_j|$ \cite{4}. The sub-coherence of $\mathbf{D}$ is defined as $\nu=\max\limits_{l}\max\limits_{i,j\neq i}|\mathbf{D}_i^T\mathbf{D}_j|, \mathbf{D}_i, \mathbf{D}_j\in \mathbf{D}[l]$.
\label{def2}
\end{def2}

\begin{def3}
For the system model (\ref{SYSTEMMODEL}), ${\rm SNR}$ is defined as $\frac{{\rm \mathbb{E}}(||\mathbf{D}\mathbf{x}||_2^2)}{{\rm \mathbb{E}}(||\mathbf{n}||_2^2)}$ \cite{12,11,13}.
\end{def3}

\begin{lemma1}
Suppose $\mu<\frac{1}{c-1}$, then $1-(c-1)\mu\leq\lambda_{\min}\leq\lambda_{\max}\leq1+(c-1)\mu$, where $c$ represents the number of the columns of $\mathbf{D}_{\mathbf{S}}$, $\mu$ is the matrix coherence of $\mathbf{D}$, $\lambda_{\min}$ and $\lambda_{\max}$ denote the minimum and maximum eigenvalue of $\mathbf{D}_{\mathbf{S}}^T\mathbf{D}_{\mathbf{S}}$ respectively \cite{3}.
\label{lemma1}
\end{lemma1}

\begin{lemma4}
Let $\mathbf{C}$ be a positive convex set, containing the origin. Suppose $h(x)\geq0$ $(x\geq0)$ be a function such that: (1) $\{x|h(x)\geq u\}=\mathbf{B}_u$ is convex for every $u$ $(0<u<\infty)$, and (2) $\int_\mathbf{C}h(x)dx<\infty$. Then, for $0\leq\beta\leq1$, $\int_{\mathbf{C}}h(x+\beta y)dx\geq \int_{\mathbf{C}}h(x+y)dx$.
\label{lemma4}
\end{lemma4}

\begin{proof}
See Appendix \ref{appendixprooflemma4}.
\end{proof}

The following corollary is derived based on Lemma \ref{lemma4}.
\begin{Corollary2}
Let $\mathbf{x}$ be a random vector with probability density function $h(x)$ $(x\geq0)$ such that: (1) $\{x|h(x)\geq u\}=\mathbf{B}_u$ is convex for every $u$ $(0\leq u<\infty)$. If $\mathbf{C}$ is a convex set, containing the origin, ${\rm P}\{\mathbf{x}+\beta y\in \mathbf{C}\}\geq {\rm P}\{\mathbf{x}+y\in \mathbf{C}\}$ for $0\leq\beta\leq1$.
\label{corollary2}
\end{Corollary2}

Based on Corollary \ref{corollary2}, the following lemma holds.
\begin{lemma5}
Let $\mathbf{x}=(\mathbf{x}_1, \mathbf{x}_2, \cdots, \mathbf{x}_c)$ be a vector of random variables having the $c$-dimensional Gamma distribution with arbitrary variances and arbitrary correlations. Then, for any positive numbers $\theta_1, \theta_2, \cdots, \theta_c$, ${\rm P}(\mathbf{x}_1\leq\theta_1, \mathbf{x}_2\leq\theta_2, \cdots, \mathbf{x}_c\leq\theta_c)\geq{\rm P}(\mathbf{x}_1\leq\theta_1)\cdot {\rm P}(\mathbf{x}_2\leq\theta_2, \cdots, \mathbf{x}_c\leq\theta_c)$.
\label{lemma5}
\end{lemma5}

The proof is omitted since it is similar to that of Theorem~1 in \cite{6}.
\section{Lower Bound of Necessary Number of Measurements}

\begin{figure}
  \centering
  \includegraphics[scale=0.5]{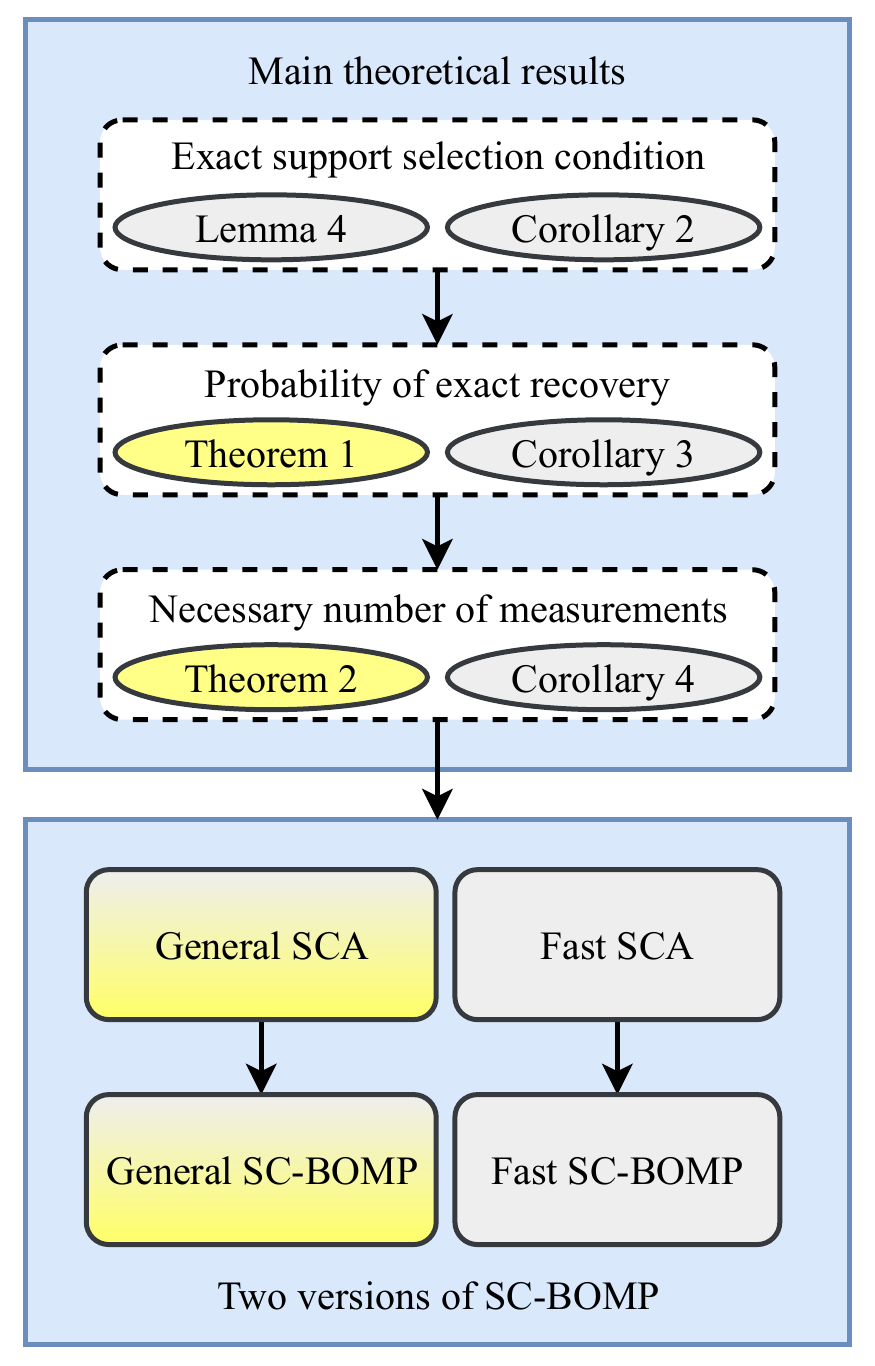}\\
  \caption{The motivation and the flow chart of this work.}\label{framework}
\end{figure}

Fig. \ref{framework} highlights the main flow from the theoretical results in Sec.~III to the proposed spectrum sensing schemes in Sec.~IV, which also demonstrates the motivation of this work that the derived theoretical results are the bases of the spectrum sensing schemes.
\subsection{Main Theoretical Results}
\label{section3a}
In this subsection, referring to the system model (\ref{SYSTEMMODEL}),  a lemma is given, which ensures that the existing BOMP algorithm exactly chooses the next support index if the previous selected indices are correct. Then, a lower bound on the probability that BOMP exactly reconstructs any $k$ block sparse signal $\mathbf{x}$ satisfying Definition \ref{def1} in $k$ iterations is provided. Finally, based on these theoretical analyses, the necessary number of measurements to guarantee that BOMP exactly recovers any $k$ block sparse signal with high probability is developed.

\subsubsection{Exact Support Selection Condition}
\begin{lemma3}
Suppose that the system model is (\ref{SYSTEMMODEL}) and $||\mathbf{n}||_2\leq\epsilon$. Let $\lambda^{\prime}_{\min}$ be the smallest singular value of $\mathbf{D}_{\mathbf{\Omega}}$. For $0\leq t\leq|\mathbf{\Omega}|-1$, assume that $\mathbf{S}^t\subseteq\mathbf{\Omega}$, $|\mathbf{S}^t|=t$ and Definition \ref{def1} holds with $\mathbf{S}=\mathbf{\Omega}\backslash \mathbf{S}^t$. Then, with $\nu<\frac{1}{d-1}$, $\mathbf{S}^{t+1}\subseteq\mathbf{\Omega}$ and $|\mathbf{S}^{t+1}|=t+1$ if the following inequality holds:
\begin{equation}\label{lemma1result}
\begin{aligned}
||\mathbf{D}^T_{\bar{\mathbf{\Omega}}}\mathbf{g}^t||_{2,\infty}<&\frac{\lambda^{\prime}_{\min}}{\sqrt{k-t}}-\frac{2\epsilon\sqrt{1+(d-1)\nu}}{\lambda^{\prime}_{\min}\sqrt{(k-t)}\min\limits_{i\in\mathbf{\Omega}}||\mathbf{x}[i]||_2},
\end{aligned}
\end{equation}
where 
\begin{equation}\label{gt}
\mathbf{g}^t=\frac{\mathbf{P}^{\bot}_{\mathbf{S}^t}\mathbf{D}_{\mathbf{\Omega}\backslash \mathbf{S}^t}\mathbf{x}_{\mathbf{\Omega}\backslash \mathbf{S}^t}}{||\mathbf{P}_{\mathbf{S}^t}^{\bot}\mathbf{D}_{\mathbf{\Omega}\backslash \mathbf{S}^t}\mathbf{x}_{\mathbf{\Omega}\backslash \mathbf{S}^t}||_2}.
\end{equation}
\label{lemma3}
\end{lemma3}

\begin{proof}
See Appendix \ref{appendixlemma3}.
\end{proof}

\begin{rmk2}
When $\mu<\frac{1}{kd-1}$, (\ref{lemma1result}) is established if the following inequality holds:
\begin{equation}\label{remark2main}
\frac{\min\limits_{i\in\mathbf{\Omega}}||\mathbf{x}[i]||_2}{\epsilon}>\frac{2\sqrt{1+(d-1)\nu}}{1-(kd-1)\mu}.
\end{equation}

Furthermore, a mutual subspace coherence $\mu_s$ as an extension of $\mu$ for matrix with block structure is defined in \cite{40}, which measures the coherence between different matrix blocks. By using $\mu_s$, an extended condition is given as follows:

When $(d-1)\nu+(k-1)d\mu_s<1$, (\ref{lemma1result}) is established if 
\begin{equation}\label{remark2mainn2}
	\frac{\min\limits_{i\in\mathbf{\Omega}}||\mathbf{x}[i]||_2}{\epsilon}>\frac{2\sqrt{1+(d-1)\nu}}{1-(d-1)\nu-(k-1)d\mu_s}.
\end{equation}
Since $\mu_s\leq\mu$ and $\nu\leq\mu$, it is obtained that $1-(d-1)\nu-(k-1)d\mu_s\geq1-(kd-1)\mu$. Thus the condition in (\ref{remark2mainn2}) is easier to hold than that in (\ref{remark2main}). 
\label{remark2}
\end{rmk2}

\begin{proof}
Based on Lemma \ref{lemma1}, $\lambda^{\prime}_{\min}\geq1-(kd-1)\mu$.
Then, let the right-side of (\ref{lemma1result}) greater than zero,  (\ref{remark2main}) holds.
By using $\mu_s$ and the similar proofs in \cite{3}, it is obtained that $\lambda^{\prime}_{\min}\geq1-(d-1)\nu-(k-1)d\mu_s$.
Finally, (\ref{remark2mainn2}) holds by using the aforementioned proofs.
\end{proof}

\begin{rmk}
By (\ref{lemma1result}) and Remark \ref{remark2},
Lemma \ref{lemma3} is easier to hold with two potential conditions: (1) the ratio of signal energy and noise is high; (3) the metrics of matrix coherence $\mu$ and sub-coherence $\nu$ are small. Based on these two assumptions, one can see intuitively that (\ref{lemma1result}) and (\ref{remark2main}) are easier to hold, and hence the performance of support selection becomes better.
\label{remark1}
\end{rmk}

By setting $\epsilon=0$, Lemma \ref{lemma3} is extended to the following corollary.
\begin{Corollary1}
Suppose the system model is $\mathbf{y}=\mathbf{D}\mathbf{x}$. Let $\lambda^{\prime}_{\min}$ be the smallest singular value of $\mathbf{D}_{\mathbf{\Omega}}$. For $0\leq t\leq|\mathbf{\Omega}|-1$, assume that $\mathbf{S}^t\subseteq\mathbf{\Omega}$, $|\mathbf{S}^t|=t$ and Definition \ref{def1} holds with $\mathbf{S}=\mathbf{\Omega}\backslash \mathbf{S}^t$. Then, $\mathbf{S}^{t+1}\subseteq\mathbf{\Omega}$ and $|\mathbf{S}^{t+1}|=t+1$ if the following inequality holds:
\begin{equation}\label{lemma1resultcorollaryversion}
\begin{aligned}
||\mathbf{D}^T_{\bar{\mathbf{\Omega}}}\mathbf{g}^t||_{2,\infty}<&\frac{\lambda^{\prime}_{\min}}{\sqrt{k-t}},
\end{aligned}
\end{equation}
where $\mathbf{g}^t$ is defined in (\ref{gt}).
\label{corollary1}
\end{Corollary1}
\subsubsection{Probability of Exact Recovery}

\begin{theoremmaindescription1}
Suppose that the system model is (\ref{SYSTEMMODEL}) and $||\mathbf{n}||_2\leq\epsilon$. Let the entries of the measurement matrix $\mathbf{D}\in \mathcal{R}^{M\times N}$ be i.i.d. $\mathcal{N}(0,\frac{1}{M})$, $\mathbf{x}$ be $k$ block sparse signal that satisfies Definition \ref{def1} and $d$ be the block length. Denote the interval:
\begin{equation}\label{interval}
\begin{aligned}
\mathcal{I}=&\Bigg(0,1-\sqrt{\frac{kd}{M}}
-\sqrt{\frac{d(c_0+1)k}{4M}}\\
&-\sqrt{\frac{d(c_0+1)k}{4M}
+\frac{2\epsilon\sqrt{1+(d-1)\nu}}{\min\limits_{i\in\mathbf{\Omega}}||\mathbf{x}[i]||_2}}\Bigg],
\end{aligned}
\end{equation}
the following inequality holds:
\begin{align}
{\rm P}(\mathbf{E})\geq&\prod_{t=1}^{k}\Bigg(1-\frac{(\frac{M\eta^2(t,\tilde{\lambda})}{d})^{\frac{d}{2}}}{\sqrt{\pi d}(\frac{M\eta^2(t,\tilde{\lambda})}{d}-1)}e^{\frac{d}{2}(1-\frac{M\eta^2(t,\tilde{\lambda})}{d})}\Bigg)^{R-k}\nonumber\\
&\times\sup_{\zeta\in\mathcal{I}}\Big(1-e^{-\frac{\zeta^2M}{2}}\Big),\label{theoremmain}
\end{align}
where $\mathbf{E}$ represents the event that BOMP reliably recovers the $k$ block sparse signal in $k$ iterations,
\begin{equation}\label{yetadefinition}
\eta(t,\tilde{\lambda})=\frac{\tilde{\lambda}}{\sqrt{t}}
-\frac{2\epsilon\sqrt{1+(d-1)\nu}}{\tilde{\lambda}\sqrt{t}\min\limits_{i\in\mathbf{\Omega}}||\mathbf{x}[i]||_2},
\end{equation}
\begin{equation}\label{eigvaluedefinition}
\tilde{\lambda}=1-\sqrt{\frac{kd}{M}}-\zeta,
\end{equation}
and $c_0$ is the positive value satisfying the inequality
\begin{equation}\label{positivesolution}
\frac{(c_0+1)^{\frac{d}{2}}}{\sqrt{\pi d}c_0}e^{-\frac{d}{2}c_0}<1.
\end{equation}
\label{theoremmaindescription1}
\end{theoremmaindescription1}

\begin{proof}
See Appendix \ref{theoremmainproof1}.
\end{proof}

\begin{rmk3}
An appropriate $c_0$ in (\ref{positivesolution}) equals $\frac{1}{2}$. The reason is given as follows. For clarity, let the left-side of (\ref{positivesolution}) be $g(x,d)=\frac{(x+1)^{\frac{d}{2}}}{\sqrt{\pi d}x}e^{-\frac{d}{2}x}$. Then, $g(\frac{1}{2}, 1)<1$. Since $\frac{\partial g(x,d)}{\partial d}<0$, $g(\frac{1}{2},d)<1$ holds for all available $d$. Furthermore, when $x>0$, $\frac{\partial g(x,d)}{\partial x}<0$. This means $g(x\geq\frac{1}{2},d)<1$. In a word, the choice of the value of $c_0$ is not fixed but limited to (\ref{positivesolution}).
\label{remark3}
\end{rmk3}

By exploiting Corollary \ref{corollary1} and setting $\epsilon=0$ in Theorem \ref{theoremmaindescription1}, the following corollary holds.
\begin{Corollary3}
Suppose that the system model is $\mathbf{y}=\mathbf{D}\mathbf{x}$. Let the entries of the measurement matrix $\mathbf{D}\in \mathcal{R}^{M\times N}$ be i.i.d. Gaussian distribution, i.e., $\mathcal{N}(0,\frac{1}{M})$, $\mathbf{x}$ be $k$ block sparse signal that satisfies Definition \ref{def1} and $d$ is a positive integer representing the block length. Denote the interval:
\begin{equation}\label{intervalcorollary22}
\begin{aligned}
\mathcal{I}=&\Bigg(0,1-\sqrt{\frac{kd}{M}}
-\sqrt{\frac{d(c_0+1)k}{M}}\Bigg],
\end{aligned}
\end{equation}
the following inequality holds:
\begin{equation}\label{theoremmaincorollary22}
\begin{aligned}
{\rm P}(\mathbf{E})\geq&\prod_{t=1}^{k}\Bigg(1-\frac{(\frac{M\eta^2(t,\tilde{\lambda})}{d})^{\frac{d}{2}}}{\sqrt{\pi d}(\frac{M\eta^2(t,\tilde{\lambda})}{d}-1)}e^{\frac{d}{2}(1-\frac{M\eta^2(t,\tilde{\lambda})}{d})}\Bigg)^{R-k}\\
&\times\sup_{\zeta\in\mathcal{I}}\big(1-e^{-\frac{\zeta^2M}{2}}\big),
\end{aligned}
\end{equation}
where $\mathbf{E}$ represents the event that BOMP reliably recovers the $k$ block sparse signal in $k$ iterations,
\begin{equation}\label{yetadefinitioncorollary22}
\eta(t,\tilde{\lambda})=\frac{\tilde{\lambda}}{\sqrt{t}},\quad\tilde{\lambda}=1-\sqrt{\frac{kd}{M}}-\zeta,
\end{equation}
and $c_0$ is the positive value satisfying (\ref{positivesolution}).
\label{corollary3}
\end{Corollary3}
\subsubsection{Necessary number of measurements}

\begin{theoremmaindescription2}
Suppose that the system model is defined in (\ref{SYSTEMMODEL}) and the atoms of $\mathbf{D}\in \mathcal{R}^{M\times N}$ are i.i.d. $\mathcal{N}(0,\frac{1}{M})$.
For $k$ block sparse signal $\mathbf{x}$ $(d>2)$ satisfying Definition \ref{def1} and $\omega\in(0,1)$, let
\begin{equation}\label{theorem222main1}
\xi=\max\bigg\{1,\frac{R-k}{\ln(N)}\times4^{2d}ke^{-\frac{d}{2}}\bigg\}
\end{equation}
and
\begin{equation}\label{theoremmain2noisecase}
\begin{aligned}
M\geq&\max\Bigg\{\bigg(\sqrt{\frac{2}{kd}}+\sqrt{\frac{1}{\ln(\frac{N}{\omega})}}+\sqrt{\frac{\alpha_2}{kd\ln(\frac{N}{\omega})}}\\
&-\sqrt{\frac{\alpha_2+\alpha_1^2\alpha_3-\alpha_2\alpha_3}{kd\ln(\frac{N}{\omega})}}\bigg)^2kd\ln\Big(\frac{N}{\omega}\Big),\\
&\Bigg(\sqrt{\frac{2}{kd}}+\sqrt{\frac{1}{\ln(\frac{N}{\omega})}}
+\frac{\beta_1\beta_2+\frac{1}{2}\beta_3}{(\min\limits_{i\in\mathbf{\Omega}}||\mathbf{x}[i]||_2-\beta_1)\sqrt{kd\ln(\frac{N}{\omega})}}\\
&+\frac{\sqrt{(\beta_1\beta_2+\frac{1}{2}\beta_3)^2+(\min\limits_{i\in\mathbf{\Omega}}||\mathbf{x}[i]||_2-\beta_1)\beta_1\beta_2^2}}{(\min\limits_{i\in\mathbf{\Omega}}||\mathbf{x}[i]||_2-\beta_1)\sqrt{kd\ln(\frac{N}{\omega})}}\Bigg)^2\\
&\times kd\ln\Big(\frac{N}{\omega}\Big)\Bigg\},
\end{aligned}
\end{equation}
where
\begin{equation}\label{ub1234=what}
\begin{aligned}
&\alpha_1=\sqrt{2\ln\Big(\frac{N}{\omega}\Big)}+\sqrt{kd}+\sqrt{\frac{d(c_0+1)k}{4}},\\
&\alpha_2=\frac{d(c_0+1)k}{4},
\alpha_3=\frac{2\epsilon \sqrt{1+(d-1)\nu}}{\min\limits_{i\in\mathbf{\Omega}}||\mathbf{x}[i]||_2},\\
&\beta_1=2\epsilon\sqrt{1+(d-1)\nu},
\beta_2=\sqrt{k}+\sqrt{2\ln\Big(\frac{N}{\omega}\Big)},\\
&\beta_3=\sqrt{d\times\bigg(\sqrt[\frac{3d}{2}+1]{\frac{\xi\ln(N)}{\omega}}+1\bigg)\times k}\times\min_{i\in\mathbf{\Omega}}||\mathbf{x}[i]||_2.
\end{aligned}
\end{equation}
Then,
\begin{equation}\label{theoremmain3}
{\rm P}(\mathbf{E})\geq1-\frac{\omega}{N}-\frac{\omega}{\sqrt{\pi d}},
\end{equation}
where $\mathbf{E}$ represents the event that BOMP reliably recovers the $k$ block sparse signal in $k$ iterations.
\label{theoremmaindescription2}
\end{theoremmaindescription2}

\begin{proof}
See Appendix \ref{theoremmainproof2}.
\end{proof}

The following corollary presents the noiseless version of Theorem \ref{theoremmaindescription2}. 
\begin{Corollary4}
Suppose that the system model is $\mathbf{y}=\mathbf{D}\mathbf{x}$ and the atoms of $\mathbf{D}\in \mathcal{R}^{M\times N}$ are i.i.d. $\mathcal{N}(0,\frac{1}{M})$. For $k$ block sparse signal $\mathbf{x}$ $(d>2)$ satisfying Definition \ref{def1} and $\omega\in(0,1)$, let $\xi=\max\bigg\{1,\frac{R-k}{\ln(N)}\times4^{2d}ke^{-\frac{d}{2}}\bigg\}$
and
\begin{equation}\label{theoremmain2noiseless1}
\begin{aligned}
M\geq&\Bigg(\sqrt{\frac{2}{kd}}+\sqrt{\frac{1}{\ln(\frac{N}{\omega})}}+\sqrt{\frac{(\sqrt[\frac{3d}{2}+1]{\frac{\xi\ln(N)}{\omega}}+1)}{\ln(\frac{N}{\omega})}}\Bigg)^2\\
&\times kd\ln\Big(\frac{N}{\omega}\Big).
\end{aligned}
\end{equation}
Then, 
\begin{equation}\label{theoremmain3}
{\rm P}(\mathbf{E})\geq1-\frac{\omega}{N}-\frac{\omega}{\sqrt{\pi d}},
\end{equation}
where $\mathbf{E}$ represents the event that BOMP reliably recovers the $k$ block sparse signal in $k$ iterations.
\label{corollary4}
\end{Corollary4}

In Theorem \ref{theoremmaindescription1}, a more reliable support selection condition than that given in \cite{14} is first developed by effectively utilizing the block structure of the measurement matrix. Specially, the support selection for one block in BOMP is more effectively formulated via the $\ell_{2,\infty}$ norm, which essentially indicates the coherence between an arbitrary matrix block and a random vector. Based on this $\ell_{2,\infty}$ form, the probability, which represents that the coherence is less than a given threshold, is higher than the result in \cite{14}. The latter one ignores the block structure in the measurement matrix. With the derived higher probability for choosing one block, the overall probability for selecting $k$ correct support blocks is thus improved.

Based on Theorem \ref{theoremmaindescription1}, Theorem \ref{theoremmaindescription2} provides a tighter bound of the necessary number of measurements than that in \cite{14} for the same probability. In the derivation of Theorem \ref{theoremmaindescription2}, an inequality relation of additive exponential function is developed by further considering the influence of block length, which finally presents an operational lower bound for the aforementioned support selection probability. This improved lower bound in turn guarantees the tightness of the necessary number of measurements.

Due to the improvement of Theorem \ref{theoremmaindescription1} and Theorem \ref{theoremmaindescription2}, our derived condition for the necessary number of measurements is thus better than the existing one. The following Remark \ref{remakr5} is presented to clearly show the superiority.

\begin{rmk5}
There exists a general form of the necessary number of measurements in previous literature \cite{2,8,14}, that is,
\begin{equation}\label{GENERALM}
M\geq Akd\ln\Big(\frac{N}{B}+C\Big)+F,
\end{equation}
where $A$, $B$, $C$ and $F$ are constants or functions. The most recent result is given in \cite{14} (Corollary 5) by asymptotic analysis:
\begin{equation}\label{recent result}
M\geq Akd\ln\Big(\frac{N}{B}\Big),
\end{equation}
where $A=2$ and $B\in[0.01,0.1]$ is an empirical constant presented in the simulation part of the paper.

By using the same asymptotic condition in \cite{14}, i.e., $k,N\rightarrow\infty$, (\ref{theoremmain2noiseless1}) becomes
\begin{equation}\label{asymptotic analysis}
\begin{aligned}
\lim_{k,N\rightarrow\infty}&\Bigg(\sqrt{\frac{2}{kd}}+\sqrt{\frac{1}{\ln(\frac{N}{\omega})}}+\sqrt{\frac{(\sqrt[\frac{3d}{2}+1]{\frac{\xi\ln(N)}{\omega}}+1)}{\ln(\frac{N}{\omega})}}\Bigg)^2\\
&<A=2.
\end{aligned}
\end{equation}
This result indicates that the condition in (\ref{theoremmain2noiseless1}) is tighter than that in \cite{14}, leading to a smaller necessary number of measurements in signal recovery.
While our result indicates the same order on $kd$ and $N$ as that in \cite{14}, it provides a tighter condition, which leads to a smaller necessary number of measurements in spectrum sensing. This tighter condition in terms of the bound on the necessary number of measurements not only sheds light on the characterization of BOMP's recoverability, as a fundamental question in the analysis of the BOMP algorithm, but also plays as a foundation for the subsequent algorithm development.
\label{remakr5}
\end{rmk5}

To better explain the physical meaning of Theorem 2, a new bound of $M$ is provided 
in the following Corollary \ref{Corollary7}.

\begin{Corollary7}
\label{Corollary7}
Suppose that the system model is defined in (1) and the atoms of $\mathbf{D}\in \mathcal{R}^{M\times N}$ are i.i.d. $\mathcal{N}(0,\frac{1}{M})$.
For $k$ block sparse signal $\mathbf{x}$ $(d>2)$ satisfying Definition \ref{def1} and $\omega\in(0,1)$, let $\xi=\max\bigg\{1,\frac{R-k}{\ln(N)}\times4^{2d}ke^{-\frac{d}{2}}\bigg\}$
and
\begin{equation}\label{theoremmain2noisecase2}
	\begin{aligned}
		M\geq&
		\Bigg(\sqrt{\frac{2}{kd}}+\sqrt{\frac{1}{\ln(\frac{N}{\omega})}}
		+\frac{2\min\limits_{i\in\mathbf{\Omega}}||\mathbf{x}[i]||_2\beta_2+\beta_3}{(\min\limits_{i\in\mathbf{\Omega}}||\mathbf{x}[i]||_2-\beta_1)\sqrt{kd\ln(\frac{N}{\omega})}}\Bigg)^2\\
		&\times kd\ln\Big(\frac{N}{\omega}\Big),
	\end{aligned}
\end{equation}
where $\beta_1$, $\beta_2$ and $\beta_3$ are defined in Theorem 2.
Then, 
\begin{equation}\label{theoremmain3}
	\mathrm{P}(\mathbf{E})\geq1-\frac{\omega}{N}-\frac{\omega}{\sqrt{\pi d}},
\end{equation}
where $\mathbf{E}$ represents the event that BOMP reliably recovers the $k$ block sparse signal in $k$ iterations.
\end{Corollary7}

\begin{proof}
As proved in Remark \ref{remark3}, $c_0$ can be set as $\frac{1}{2}$. Then, the first term in (\ref{theoremmain2noisecase}) satisfies:
\begin{equation}\label{corollary 51}
	\begin{aligned}
		&\bigg(\sqrt{\frac{2}{kd}}+\sqrt{\frac{1}{\ln(\frac{N}{\omega})}}+\sqrt{\frac{\alpha_2}{kd\ln(\frac{N}{\omega})}}
		-\sqrt{\frac{\alpha_2+\alpha_1^2\alpha_3-\alpha_2\alpha_3}{kd\ln(\frac{N}{\omega})}}\bigg)^2\\
		&\times kd\ln\Big(\frac{N}{\omega}\Big)
		\leq\bigg(\sqrt{\frac{2}{kd}}+\sqrt{\frac{1}{\ln(\frac{N}{\omega})}}\bigg)^2
		kd\ln\Big(\frac{N}{\omega}\Big).
	\end{aligned}
\end{equation}
Meanwhile, by letting $\min\limits_{i\in\mathbf{\Omega}}||\mathbf{x}[i]||_2>\beta_1$, the last item in parentheses of the second item in (\ref{theoremmain2noisecase}) satisfies:
\begin{equation}\label{corollary 52}
	\begin{aligned}
		&\frac{\sqrt{(\beta_1\beta_2+\frac{1}{2}\beta_3)^2+(\min\limits_{i\in\mathbf{\Omega}}||\mathbf{x}[i]||_2-\beta_1)\beta_1\beta_2^2}}{(\min\limits_{i\in\mathbf{\Omega}}||\mathbf{x}[i]||_2-\beta_1)\sqrt{kd\ln(\frac{N}{\omega})}}\\
		\leq& \frac{\min\limits_{i\in\mathbf{\Omega}}||\mathbf{x}[i]||_2\beta_2+\frac{1}{2}\beta_3}{(\min\limits_{i\in\mathbf{\Omega}}||\mathbf{x}[i]||_2-\beta_1)\sqrt{kd\ln(\frac{N}{\omega})}}.
	\end{aligned}
\end{equation}
\\Integrating the above derivations, and using the similar procedures in the proof of Theorem 2, the proof of Corollary 5 is completed.
\end{proof}

The lower bound in Corollary 5 is more intuitive than that in Theorem 2. Based on (7), there are two observations: 1) the lower bound of $M$ decreases if  $\min\limits_{i\in\mathbf{\Omega}}||\mathbf{x}[i]||_2$ increases; 2) the lower bound of $M$ becomes smaller if the noise constraint $\epsilon$ is reduced. That is, the increase of useful signal power or the decrease of noise power makes it easier to recover the spectrum signal.

\begin{Corollary5}
Suppose that the system model is (\ref{SYSTEMMODEL}) and $||\mathbf{n}||_2\leq\epsilon$. Let the entries of the measurement matrix $\mathbf{D}\in \mathcal{R}^{M\times N}$ be i.i.d. Gaussian distribution, i.e., $\mathcal{N}(0,\frac{1}{M})$, $\mathbf{x}$ be $k$ block sparse signal that satisfies Definition \ref{def1} and $d$ be the block length.

For $\vartheta\in(0,\frac{1}{\sqrt{\pi d}}]$, let
\begin{equation}\label{omegavalue}
\omega=\frac{\vartheta N\sqrt{\pi d}}{\sqrt{\pi d}+N},
\end{equation}
and $M$ satisfy (\ref{theoremmain2noisecase}). Then,
\begin{equation}\label{psomega}
{\rm P}(\mathbf{E})\geq1-\vartheta,
\end{equation}
where $\mathbf{E}$ represents the event that BOMP reliably recovers the $k$ block sparse signal in $k$ iterations.
\label{corollary5}
\end{Corollary5}

\begin{proof}
Since $\vartheta\leq \frac{1}{\sqrt{\pi d}}<\frac{1}{N}+\frac{1}{\sqrt{\pi d}}$, then $\omega<1$, which satisfies the assumption of that in Theorem \ref{theoremmaindescription2}, i.e., $\omega\in(0,1)$. Meanwhile,
\begin{equation}\label{appendixproofofaorollary52}
\frac{\omega}{N}+\frac{\omega}{\sqrt{\pi d}}=\vartheta.
\end{equation}

Finally, (\ref{psomega}) holds and the corollary is concluded.
\end{proof}

Corollary \ref{corollary5} also applies to Corollary \ref{corollary4} which provides the necessary number of measurements under noiseless case.

The theoretical analyses extend the basic idea of analyzing the support selection conditions in \cite{14,20}. However, different from \cite{14,20}, the developed results in this paper consider the block structure and utilize the block MIP to bound the mixed matrix norm. They provide considerable improvement and further reduce the computing resources required for reliable recovery. Another contribution is that this paper is the first to consider the noise impact on the bounds of the necessary number of measurements. The results for the noisy scenarios offer effective guideline for the implementation of practical CSS.

\section{Sampling-Controlled Block Orthogonal Matching
Pursuit}
In this section, the sampling-controlled algorithm (SCA) is first provided based on the theoretical results in Section \ref{section3a}. SCA is a preprocessing procedure on the given sparse signals, which determines the decent number of measurements, given the block sparsity, block length and signal energy. In practical spectrum sensing applications, SCA allows to dynamically guide the measurement matrix based on the calculated number of measurements, resulting in an efficient spectrum sensing scheme.
Further, the integration of SCA and BOMP leads us to design a novel sampling-controlled block orthogonal matching pursuit (SC-BOMP) as well as its fast implementation version.

\subsection{The Proposed SC-BOMP Schemes}

\subsubsection{The General SC-BOMP}

The general SCA is given in Algorithm~\ref{SCA} by exploiting Theorem \ref{theoremmaindescription2} and Corollary~\ref{corollary4}.
In step 1, the algorithm obtains a lower bound of the number of measurements $M_{\min}$ by (\ref{theoremmain2noiseless1}) in Corollary~\ref{corollary4}, which is the minimum number of measurements but not applicable to the noisy case. To explain why this $M_{\min}$ is necessary for the general SCA, two points are clarified:  \textbf{\emph{(a) the lower bound of $M$ in (\ref{theoremmain2noiseless1}) is independent to the sub-coherence $\nu$ of the measurement matrix; (b) by calculation, the lower bound of $M$ in (\ref{theoremmain2noisecase}) decreases with the decrease of $\nu$.}}

Generally speaking, the sub-coherence $\nu$ of the measurement matrix whose entries are of static distribution changes qualitatively but not quantitatively with the variation of $M$, that is, $\nu$ increases with the decrease of $M$, vice versa.
Therefore, in step 3, $M_{\min}$ produces the largest sub coherence $\nu$, which can be seen as the worst $\nu$, causing the largest $M_{\max}$ with respect to sub coherence $\nu$ in (\ref{theoremmain2noisecase}). The bound of $M_{opt}$ is given as follows:
\begin{equation}\label{moptimal}
M_{\min}<M_{opt}\leq M_{\max}.
\end{equation}

The first inequality in (\ref{moptimal}) is because $M_{opt}$ can not be $M_{\min}$ due to the existence of noise in practical applications.
In SCA, to be conservative, choose $M_{opt}\approx M_{\max}$.
Since the decent number of measurements should be small but not the largest, there exists gap between $M_{\max}$ and $M_{opt}$. Fortunately, this gap is negligible as long as the minimal energy of the signal block is sufficient large or the energy of the noise is small.

In steps 4 and 5 of the SCA, there exists an exponential mapping indicator $e^{(\cdot)}$, which is widely used in many applications \cite{32,30,31,29}, such as face recognition and hyperspectral image classification. In step 4, this indicator can map $\epsilon$ to lower level values for the compensation of theoretical scaling in Theorem \ref{theoremmaindescription2} and Corollary~\ref{corollary4}, which is the common shortcoming of the necessary number of measurements \cite{14}. In step 5, the first term of the indicator, i.e., $e^{-\varepsilon_210^{\frac{{\rm SNR}}{10}}}$, is the similar scaling mapping to that of in step 4. The second term in the indicator in step 5, i.e., $e^{-\varepsilon_2\varepsilon_3}$, guarantees that the trimmed $M_{opt-trim}$ cannot be infinitely close to 0.

\begin{figure*}
  \centering
  \includegraphics[scale=0.6]{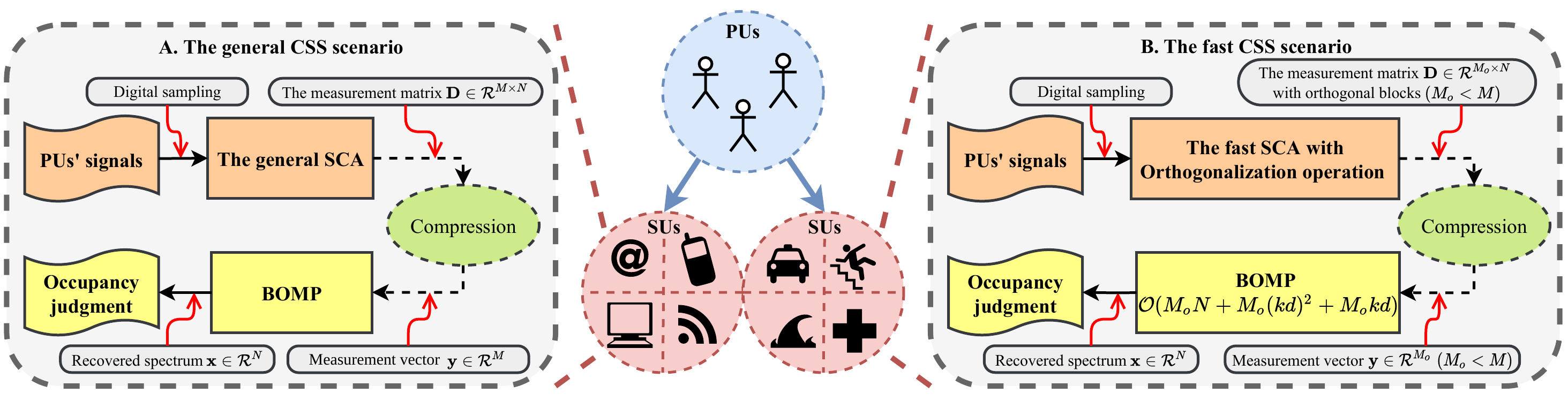}\\
  \caption{The flow chart of the proposed spectrum sensing schemes.}\label{SC-BOMPlast}
\end{figure*}

The holistic general SC-BOMP scheme is shown in Fig. \ref{SC-BOMPlast} (A).
\begin{algorithm}
	\renewcommand{\algorithmicrequire}{\textbf{Input:}}
	\renewcommand{\algorithmicensure}{\textbf{Output:}}
	\caption{General SCA}
	\label{alg:33}
	\begin{algorithmic}[1]
		\REQUIRE Sparse signal $\mathbf{x}$, block sparsity level $k$, block length $d$, $c_0$, $\varepsilon_1$, $\varepsilon_2$, $\varepsilon_3$, SNR and the distribution of each atom in the measurement matrix $\mathbf{D}$
		\ENSURE The measurement matrix $\mathbf{D}$
        \STATE Calculate $M_{\min}$ by using the lower bound in (\ref{theoremmain2noiseless1})
        \STATE Set $M_{\min}=\lceil M_{\min}\rceil$
		\STATE Calculate $\nu_{\max}$ by using Definition \ref{def2} with the help of $M_{\min}$
		\STATE Calculate $M_{opt}$ by using Theorem \ref{theoremmaindescription2} with $\nu_{\max}$ and $\epsilon=\varepsilon_1 e^{-10^{\frac{{\rm SNR}}{10}}}$
        \STATE Calculate $M_{opt-trim}=(e^{-\varepsilon_210^{\frac{{\rm SNR}}{10}}}+e^{-\varepsilon_2\varepsilon_3})\times M_{opt}$
        \STATE Set $M_{opt-trim}=\lceil M_{opt-trim}\rceil$
        \STATE Generate the measurement matrix $\mathbf{D}$ with $M_{opt}$ and the corresponding distribution
		\STATE \textbf{return} $\mathbf{D}$
	\end{algorithmic}
\label{SCA}
\end{algorithm}

\subsubsection{The Fast SC-BOMP}

In some situations, the speed of detection is more crucial than its accuracy, under which a fast implementation of SC-BOMP is preferred.

Apply block orthogonality, i.e., $\nu=0$, to the measurement matrix. Under this condition, (\ref{lemma1result}) and (\ref{remark2main}) are more easier to hold, leading to better performance of support selection. Accordingly, the necessary number of measurements becomes lower when the recovery probability is the same as before, resulting in faster sensing speed. The following corollary is presented based on the block orthogonality.
\begin{Corollary6}
Suppose that the system model is defined in (\ref{SYSTEMMODEL}) and the atoms of $\mathbf{D}\in \mathcal{R}^{M\times N}$ satisfy $\mathcal{N}(0,\frac{1}{M})$.
For $k$ block sparse signal $\mathbf{x}$ $(d>2)$ satisfying Definition \ref{def1} and $\omega\in(0,1)$, let $\xi=\max\bigg\{1,\frac{R-k}{\ln(N)}\times4^{2d}ke^{-\frac{d}{2}}\bigg\}$
and $M$ bounded by (\ref{theoremmain2noisecase}) with
\begin{equation}\label{corollary6ub1234=what}
\begin{aligned}
&\alpha_1=\sqrt{2\ln\Big(\frac{N}{\omega}\Big)}+\sqrt{kd}+\sqrt{\frac{d(c_0+1)k}{4}},\\
&\alpha_2=\frac{d(c_0+1)k}{4},\alpha_3=\frac{2\epsilon}{\min\limits_{i\in\mathbf{\Omega}}||\mathbf{x}[i]||_2},\\
&\beta_1=2\epsilon,\beta_2=\sqrt{k}+\sqrt{2\ln\Big(\frac{N}{\omega}\Big)},\\
&\beta_3=\sqrt{d\times\bigg(\sqrt[\frac{3d}{2}+1]{\frac{\xi\ln(N)}{\omega}}+1\bigg)\times k}\times\min_{i\in\mathbf{\Omega}}||\mathbf{x}[i]||_2.
\end{aligned}
\end{equation}
Then, 
\begin{equation}\label{corollary6theoremmain3}
{\rm P}(\mathbf{E})\geq1-\frac{\omega}{N}-\frac{\omega}{\sqrt{\pi d}},
\end{equation}
where $\mathbf{E}$ represents the event that BOMP reliably recovers the $k$ block sparse signal in $k$ iterations.
\label{corollary6}
\end{Corollary6}

By using Corollaries \ref{corollary4} and \ref{corollary6}, the fast SCA is given in Algorithm \ref{fastSCA}. Finally, combining the fast SCA and BOMP algorithm, the fast SC-BOMP scheme is given in Fig.~\ref{SC-BOMPlast} (B). It can be seen that the complexity of BOMP using fast SCA, i.e., $\mathcal{O}(M_oN+M_o(kd)^2+M_okd)$, is lower than that of the one using general SCA at the cost of only a slight accuracy loss, where $M_o$ laconically represents the decent number of measurements, i.e., $M_{opt-trim}$, obtained from the fast SCA.

In a nutshell, the proposal of SC-BOMP schemes relies on the following promotions of the theoretical results: 1) 
this work first derives the necessary number of measurements that is reduced than the existing one, which thus reduces the gap between the empirical bound and the theoretical results; 2) then, the aforementioned gap is further improved based on the exponential mapping indicator, and the general SC-BOMP is proposed based on these theoretical results; 3) finally, based on the block orthogonality, a tighter bound than the previous two is derived, which is the theoretical foundation of the fast SC-BOMP. Compared with the existing results, our results provide an effective guidance for practical designs. They formulate the real-time SC-BOMP schemes whose numbers of measurements are just sufficiently enough for reliable CSS without the waste of computing resources.

\begin{algorithm}
	\renewcommand{\algorithmicrequire}{\textbf{Input:}}
	\renewcommand{\algorithmicensure}{\textbf{Output:}}
	\caption{Fast SCA}
	\label{alg:33}
	\begin{algorithmic}[1]
		\REQUIRE Sparse signal $\mathbf{x}$, block sparsity level $k$, block length $d$, $c_0$, $\varepsilon_1$, $\varepsilon_2$, $\varepsilon_3$, SNR and the distribution of each atom in the measurement matrix $\mathbf{D}$
		\ENSURE The measurement matrix $\mathbf{D}$
		\STATE Calculate $M_{opt}$ by using Corollary \ref{corollary6} with $\epsilon=\varepsilon_1 e^{-10^{\frac{{\rm SNR}}{10}}}$
		\STATE Calculate $M_{opt-trim}=(e^{-\varepsilon_210^{\frac{{\rm SNR}}{10}}}+e^{-\varepsilon_2\varepsilon_3})\times M_{opt}$
		\STATE Set $M_{opt-trim}=\lceil M_{opt-trim}\rceil$
		\STATE Generate the measurement matrix $\mathbf{D}$ with $M_{opt-trim}$ and the corresponding distribution
		\STATE \textbf{return} $\mathbf{D}$
	\end{algorithmic}
	\label{fastSCA}
\end{algorithm}

\section{Simulation Results}

In this section, our theoretical results are first compared with the existing bounds. Then, the proposed SC-BOMP schemes are applied to CSS.

\subsection{Comparison of theoretical results}

Since the classical bounds of necessary number of measurements are given under noiseless case, this work compares them with our derived noiseless bound. Specifically, the simulation results of Corollary~\ref{corollary4} based on Corollary~\ref{corollary5} called ``New bound'' in this paper are given, paralleled with the well known result of Corollary 7 in \cite{20} called ``Existing bound~1'', i.e., $M\geq4kd\ln(\frac{2N}{\omega_1})$, and the most recent result of Corollary 5 in \cite{14} called ``Existing bound~2'', i.e., $M\geq2kd\ln(\frac{N}{\omega_2})$.
The calculations of $\omega_1$ and $\omega_2$ are given in Eqn. 53 \cite{14} and Eqn. 16 \cite{14}, respectively.
Set $\vartheta=0.1:-0.01:0.01$ to guarantee that ${\rm P}(\mathbf{E})$ in (\ref{psomega}) is not smaller than $1-\vartheta$. The $\omega$ in (\ref{theoremmain2noiseless1}) is set the same as (\ref{omegavalue}), i.e., $\omega=\frac{\vartheta N\sqrt{\pi d}}{\sqrt{\pi d}+N}$. The block sparse signal is binary. Specially, the signal to be recovered has one entries on the randomly chosen support set $\mathbf{\Omega}$, where the positions of nonzero blocks of the $k$ block sparse signal are selected at random among all $R$ locations and the non support elements are set to be zero.

\begin{figure*}
	\centering
	\subfigure[]{\label{fig:subfig:a}
		\includegraphics[width=0.45\linewidth]{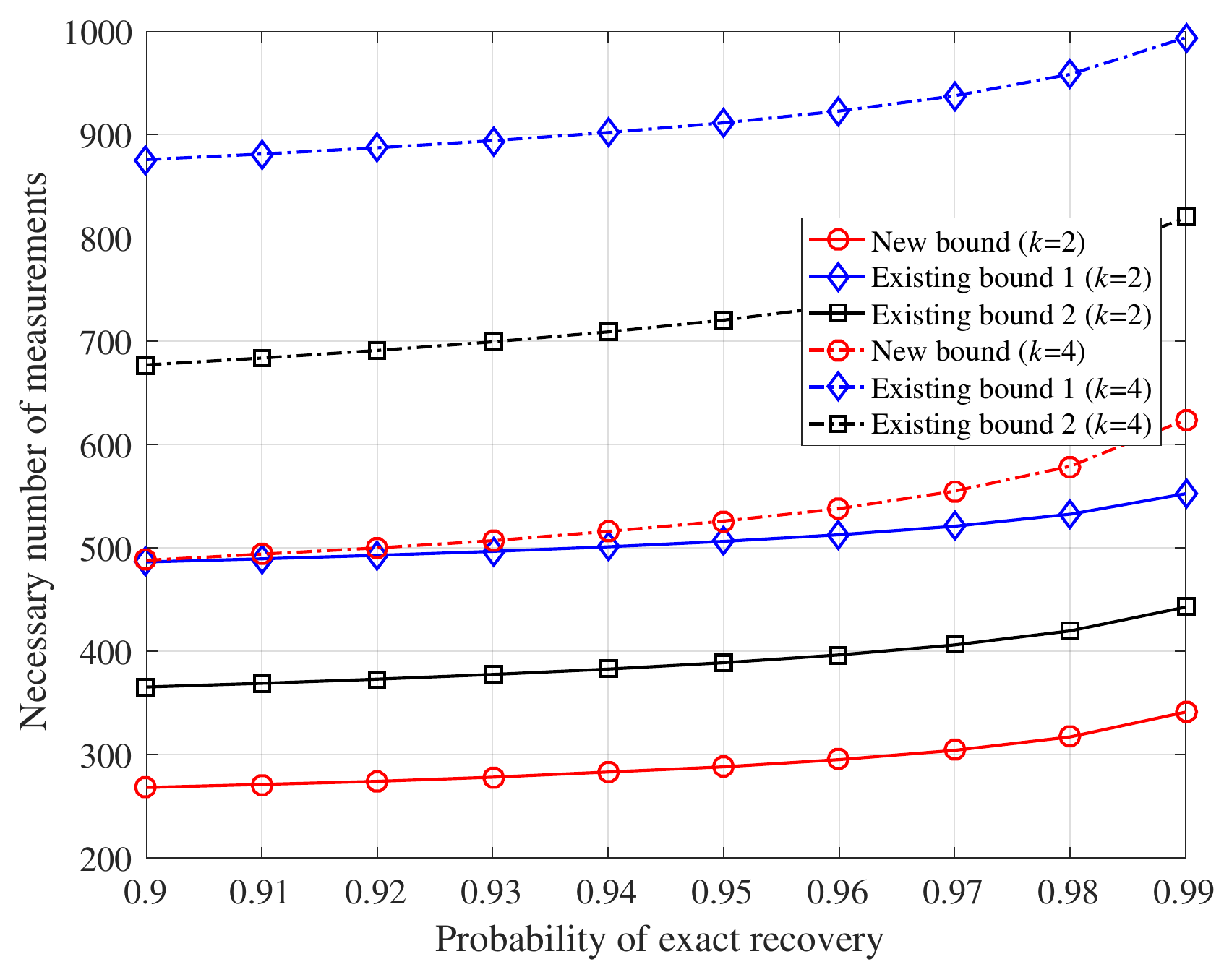}}
	\hspace{0.01\linewidth}
	\subfigure[]{\label{fig:subfig:b}
		\includegraphics[width=0.45\linewidth]{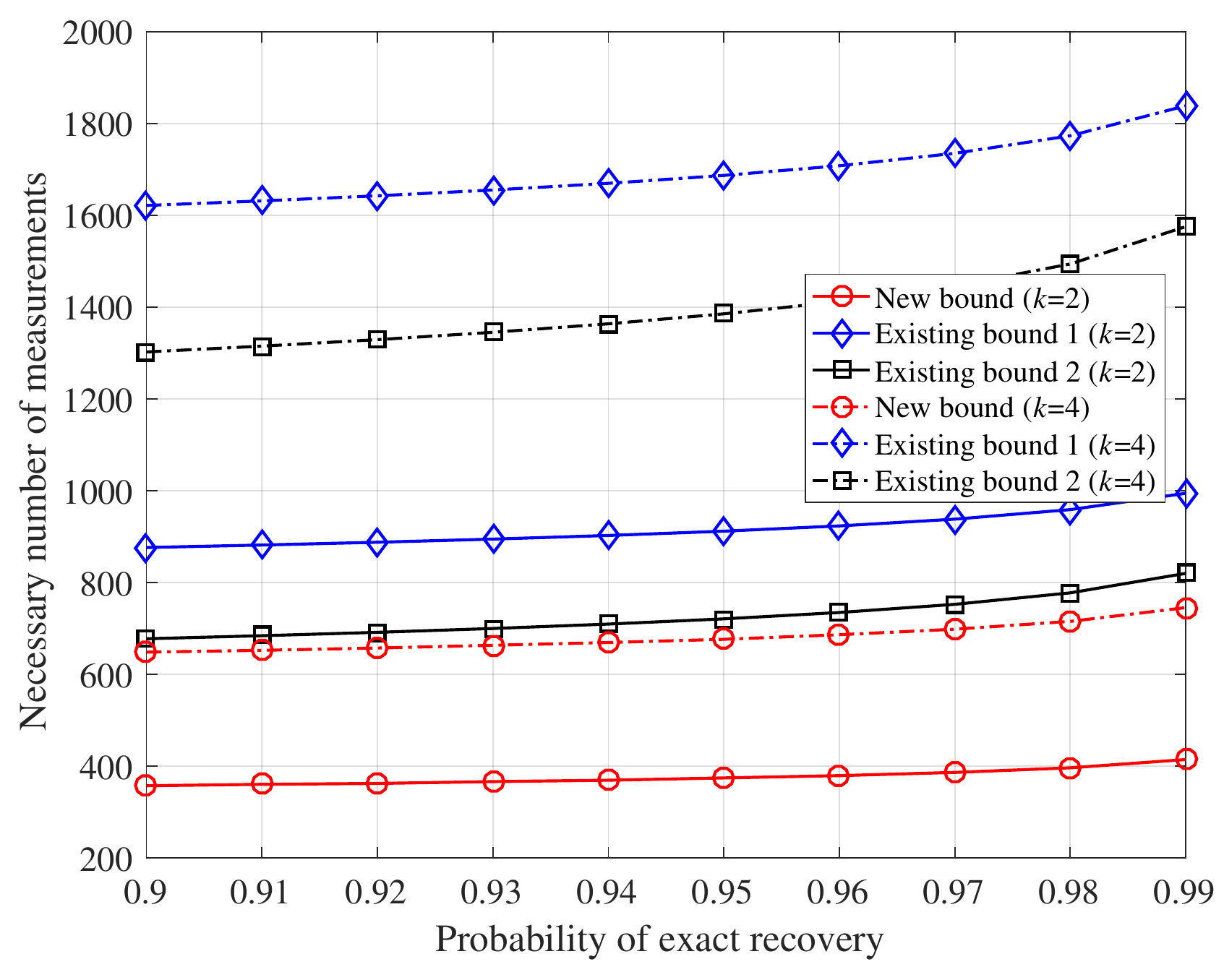}}
	\caption{Lower bounds on the necessary number of measurements $M$ for recovering $k$ block sparse signals in dimension $N=1024$ with (a) $d=4$; (b) $d=8$.}
	\label{necessary1}
\end{figure*}

The results are given in Fig. \ref{necessary1} by varying different block length $d$.
It is observed that the different ``New bounds'' are smaller than the ``Existing bounds'' which indicates that Corollary~\ref{corollary4} is better than Corollary 7 in \cite{20} and Corollary 5 in \cite{14} on describing the necessary number of measurements that guarantees the probability of exact recovery ${\rm P}(\mathbf{E})$ is not smaller than a given probability. When the block length $d$ increases with the fixed total sparsity $kd$, i.e., the ``New bound $(k=4)$'' in Fig. \ref{necessary1} (a) and the ``New bound $(k=2)$'' in Fig. \ref{necessary1} (b), the necessary number of measurements for the probability of exact recovery not being smaller than a given probability becomes better, which indicates that the gap between the empirical and theoretical number of measurements is reduced.
In addition, by comparing Figs. \ref{necessary1} (a) and (b), it can be inferred that when the total sparsity is fixed, the larger block length makes the performance of the BOMP better, because the larger block length makes the lower bound of the number of measurements required by the BOMP smaller.

\subsection{SC-BOMP Schemes for CSS}

This subsection applies our proposed general and fast SC-BOMP schemes to CSS application.

Consider a wideband CRN with the spectrum length $N=1024$. The distribution of the entry of the spectrum $\mathbf{x}$ is given as follows \cite{27}:
\begin{equation}\label{spectrumxdis}
\begin{aligned}
\mathbf{x}_i\left\{\begin{matrix}\sim \mathcal{N}(1,0.01),&i\in\mathbf{\Omega},\\
=0,&i\notin\mathbf{\Omega}.
\end{matrix}\right.
\end{aligned}
\end{equation}
The measurement matrix is Gaussian matrix with each entry satisfying i.i.d. $\mathcal{N}(0,\frac{1}{M})$, where $M$ is designed by SCAs. The parameters $\varepsilon$ and $c_0$ are set as 0.1 and 0.5 respectively. $10,000$ block sparse spectrum is generated first. Then, the general SCA and the fast SCA are used to design the measurement matrix for the SC-BOMP schemes. For the conventional BOMP scheme, the determined matrix with fixed compression rate is used.
The following two metrics are employed to evaluate different schemes.

The probability of exact detection $P_d$ is exploited as the metric to measure the recovery performance. Given the true state $\mathbf{x}\in\{0,1\}^{N}$ of the target wideband spectrum, $P_d$ is defined as follows:
\begin{equation}\label{PD}
P_d = \frac{\mathbf{x}^T(\mathbf{x}==\hat{\mathbf{x}})}{\mathbf{x}^T(\mathbf{x}==\hat{\mathbf{x}})+\mathbf{x}^T(\mathbf{x}\neq \hat{\mathbf{x}})},
\end{equation}
where $\hat{\mathbf{x}}$ is the estimated spectrum, $\mathbf{x}==\hat{\mathbf{x}}$ and $\mathbf{x}\neq \hat{\mathbf{x}}$ represent the logical operations of ``and'' and ``exclusive or'', respectively.

The running time is used to measure their efficiency. The SCA is an offline operation, thus its running time is not considered in the simulation. Our default testing environment is Matlab R2016a on a desktop computer with 2.90GHz Intel Core i7-10700 CPU.

\begin{figure*}
\centering
\subfigure[]{\label{fig:subfig:a}
\includegraphics[width=0.45\linewidth]{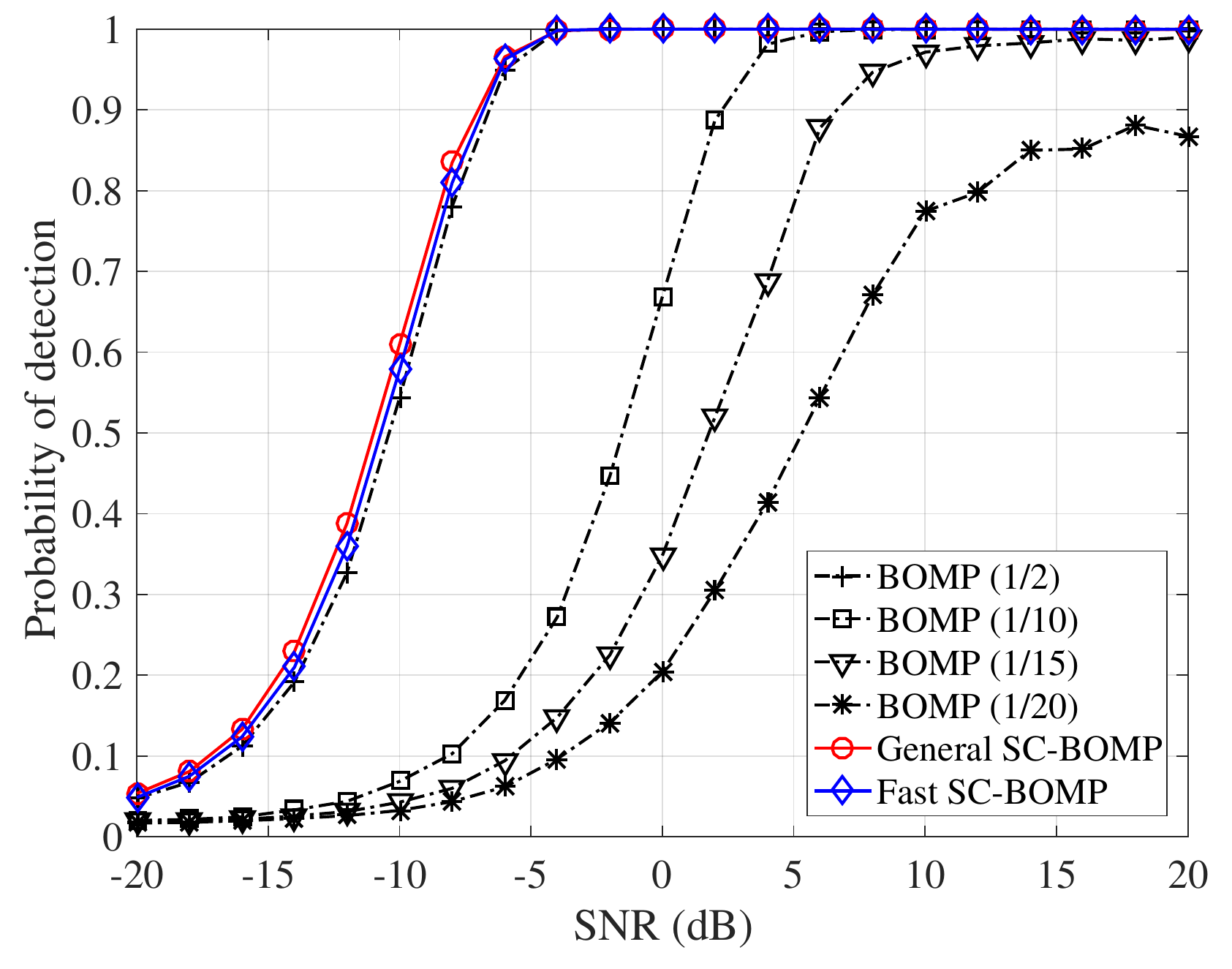}}
\hspace{0.01\linewidth}
\subfigure[]{\label{fig:subfig:b}
\includegraphics[width=0.45\linewidth]{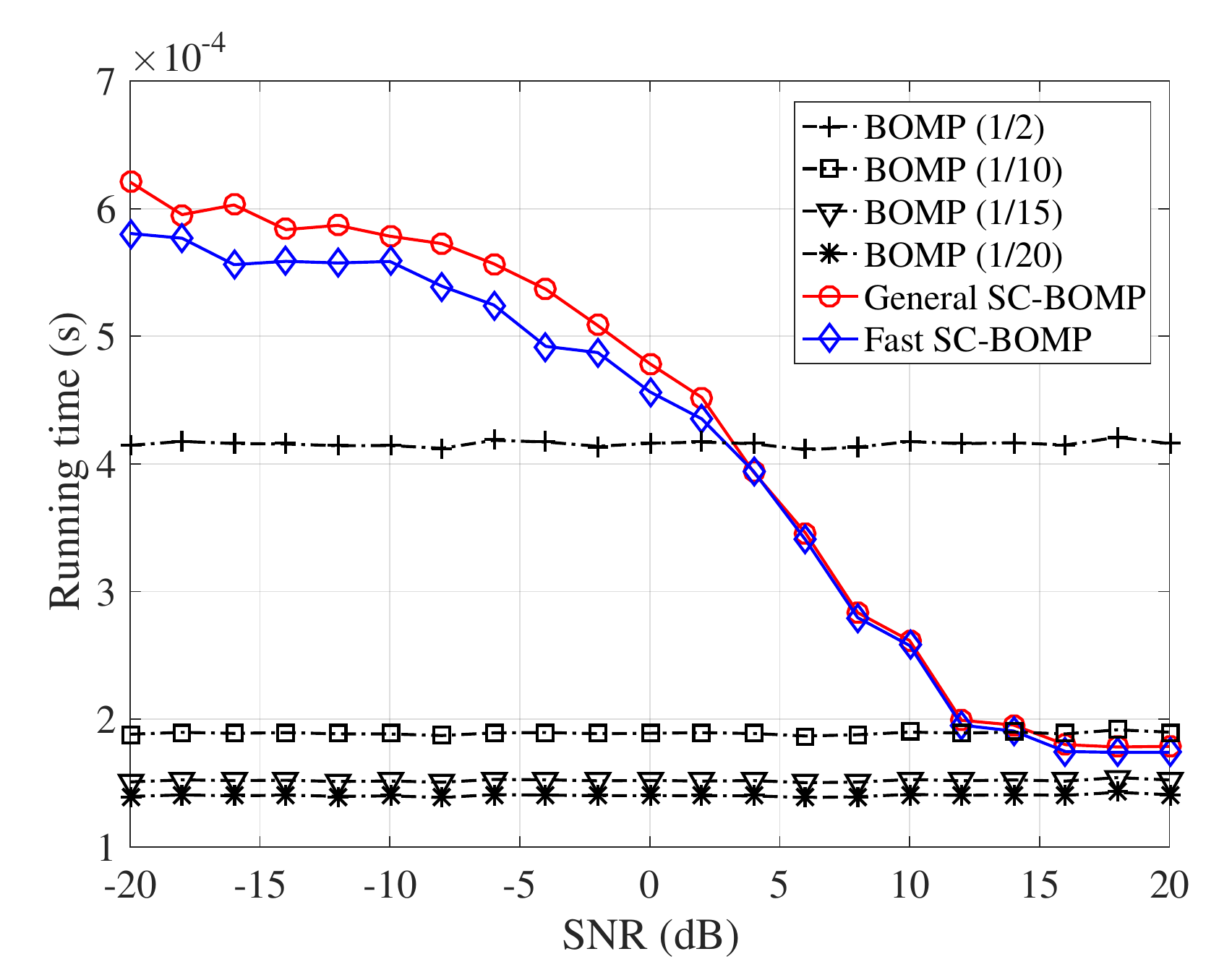}}
\hspace{0.01\linewidth}
\subfigure[]{\label{fig:subfig:c}
\includegraphics[width=0.45\linewidth]{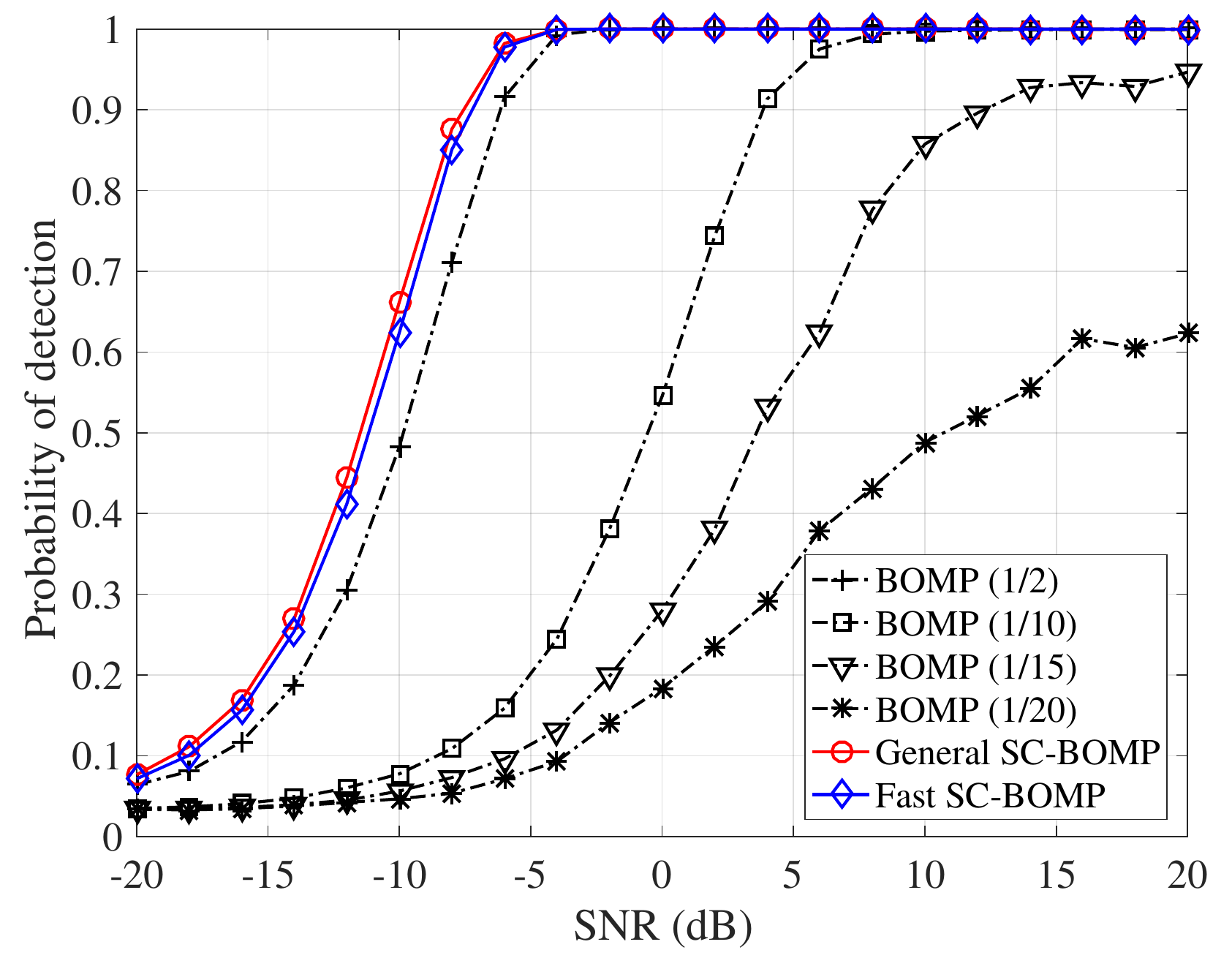}}
\hspace{0.01\linewidth}
\subfigure[]{\label{fig:subfig:d}
\includegraphics[width=0.45\linewidth]{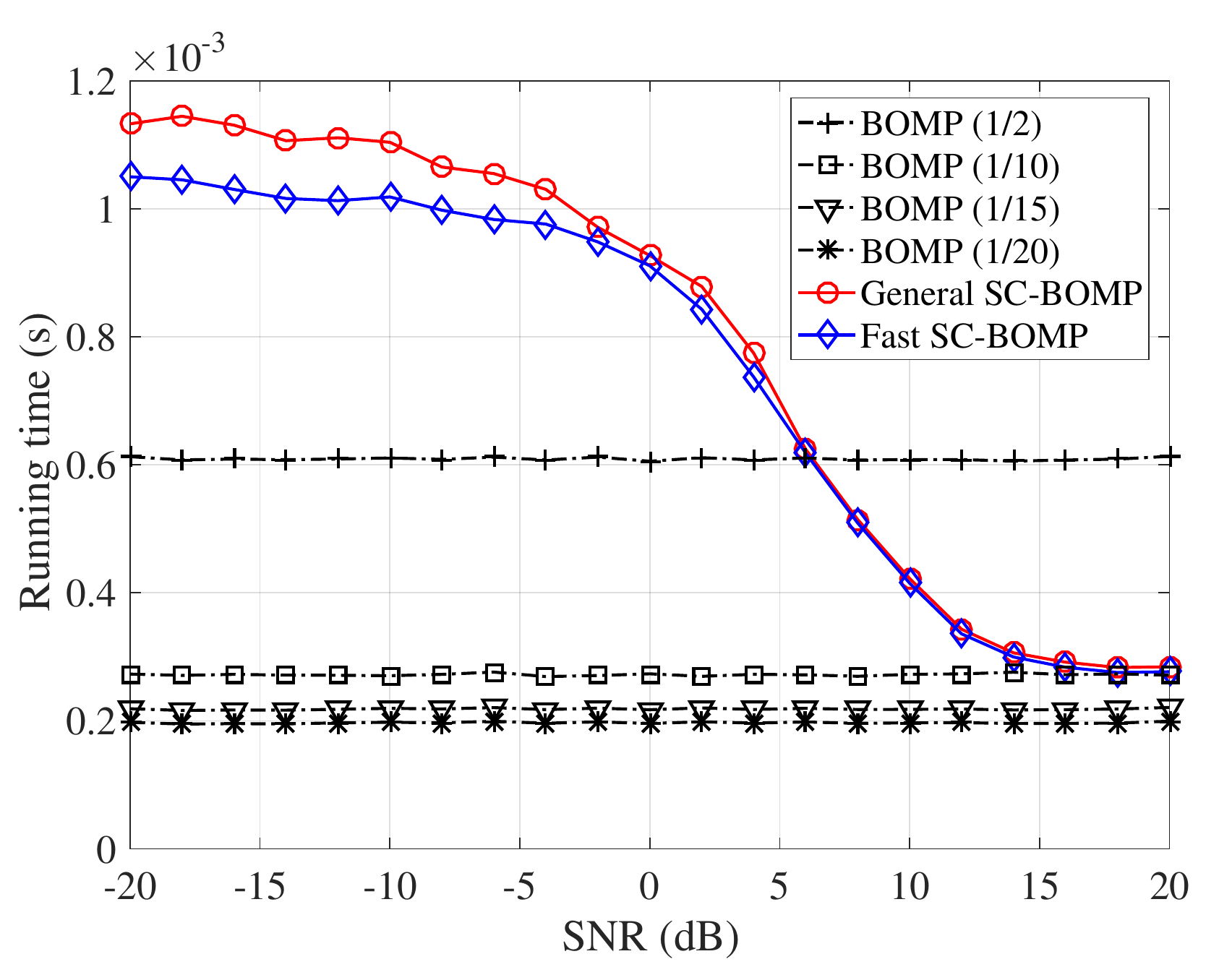}}
\hspace{0.01\linewidth}
\subfigure[]{\label{fig:subfig:e}
\includegraphics[width=0.45\linewidth]{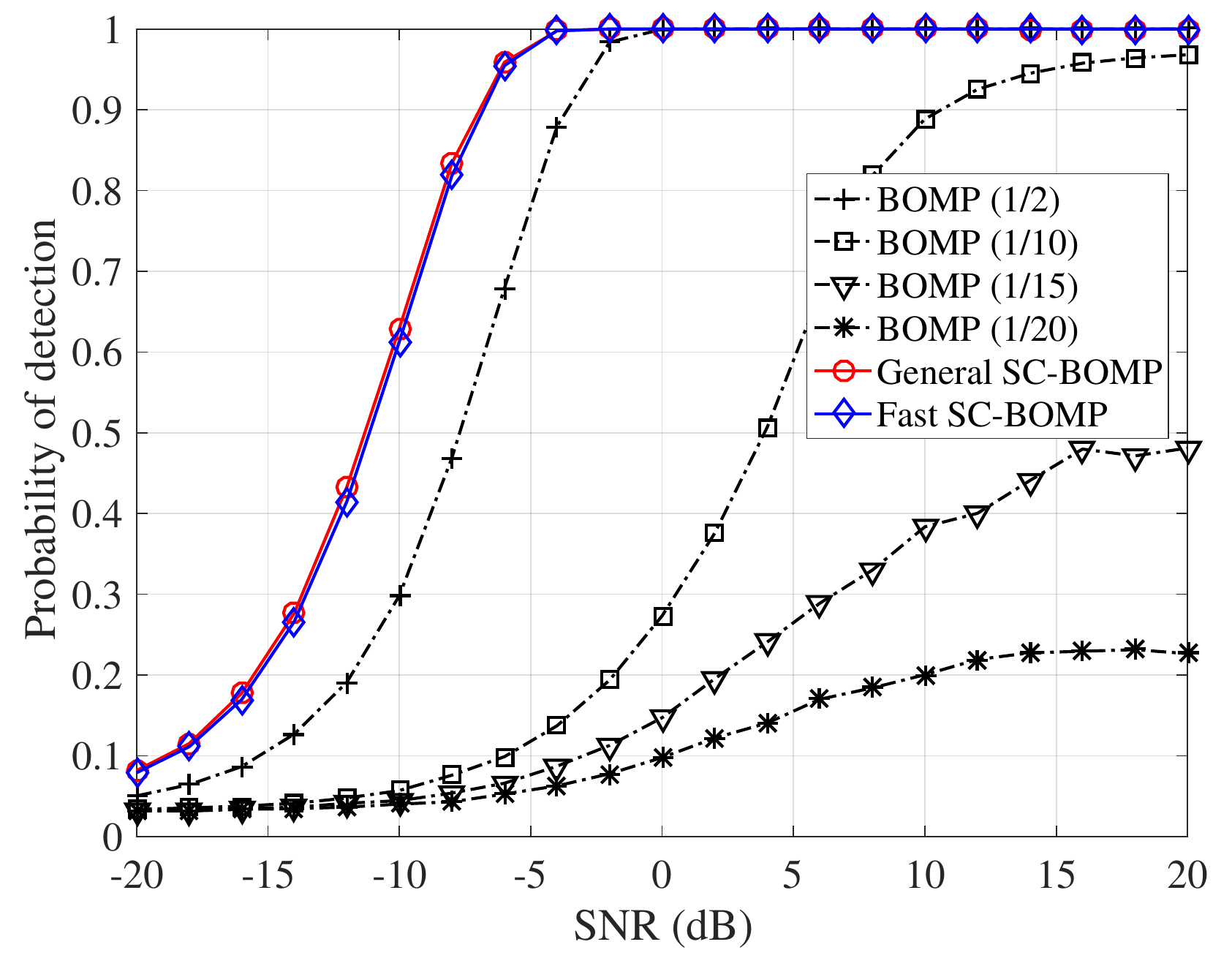}}
\hspace{0.01\linewidth}
\subfigure[]{\label{fig:subfig:f}
\includegraphics[width=0.45\linewidth]{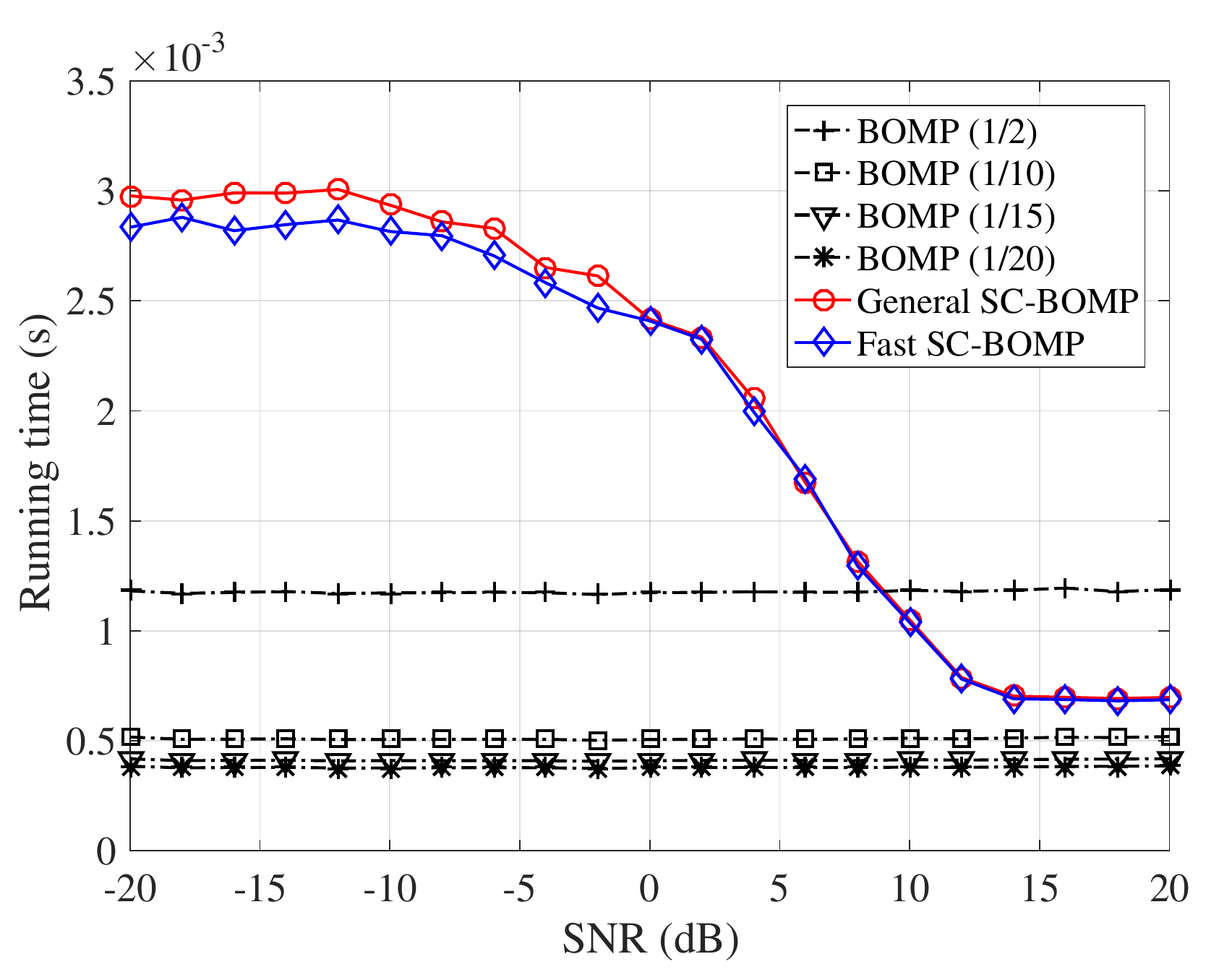}}
\hspace{0.01\linewidth}
\caption{Results of CSS with $N=1024$. (a) and (b) are with $k=4$ and $d=4$; (c) and (d) are with $k=4$ and $d=8$; (e) and (f) are with $k=8$ and $d=4$.}
\label{CSSsimulationres}
\end{figure*}

The simulation results are given in Fig. \ref{CSSsimulationres}.
It is observed that the sensing performance of our proposed SC-BOMP schemes is slightly better that that of the scheme using BOMP with the 1/2 compression rate, called BOMP (1/2) this paper and so on. Meanwhile, the running time of our SC-BOMP schemes are worse than that of BOMP (1/2) when the SNR is less than 0 dB. However, as the SNR increases, our proposed SC-BOMP schemes dynamically reduce the number of measurements, resulting in the effective complexity reduction.

It can be inferred from Fig. \ref{CSSsimulationres} (a) that BOMP (1/15) cannot exactly recovers the spectrum when SNR is lower than 20 dB. Then, in Fig. \ref{CSSsimulationres} (b), the running time of the SC-BOMP schemes are between those of BOMP (1/10) and BOMP (1/15) when SNR is large enough for exact recovery. These results indicate that the number of measurements of the SC-BOMP schemes have reduced to appreciably small values. That is, the sensing performance of the SC-BOMP schemes cannot be exact if the numbers of measurements become smaller, while the numbers of measurements are excessive if they become larger.

In Figs. \ref{CSSsimulationres} (c) and \ref{CSSsimulationres} (d), the block length is doubled to 8. The probabilities of detection of BOMP (1/2) get worse than those in Fig. \ref{CSSsimulationres} (a). However, the probabilities of detection of our proposed SC-BOMP schemes become better, leading to larger performance gain compared with BOMP (1/2). In Fig. \ref{CSSsimulationres} (d), the running time of all the schemes are worse than those in Fig. \ref{CSSsimulationres} (b). It can be seen that our SC-BOMP schemes still exhibit strong ability in reducing sensing complexity.

In Figs. \ref{CSSsimulationres} (e) and \ref{CSSsimulationres} (f), the block sparsity is doubled to 8. Our proposed SC-BOMP schemes maintain the reliable performance. Except SC-BOMP schemes, the performance of other algorithms decrease abominably. These results reveal the strong stability of the SC-BOMP schemes while facing various spectrum.

Furthermore, it can be seen that the performance of the fast SC-BOMP scheme is worse than that of the general SC-BOMP scheme, while the running time of the fast one is shorter than the general one in low SNR condition. These results indicate that the fast SC-BOMP scheme is more suitable for the scenarios requiring rapid spectrum sensing.

\begin{figure*}
	\centering
	\subfigure[]{\label{fig:subfig:a}
		\includegraphics[width=0.45\linewidth]{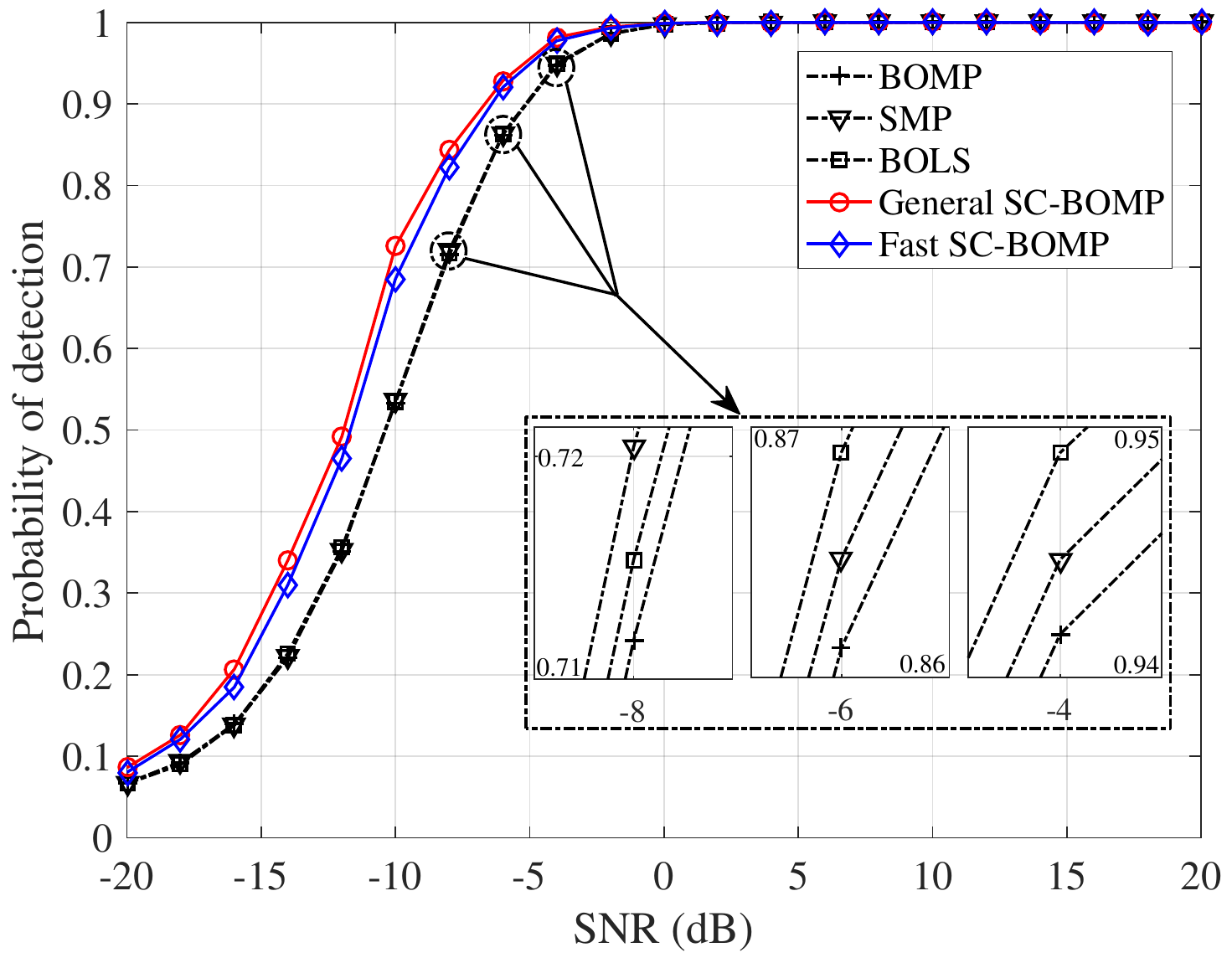}}
	\hspace{0.01\linewidth}
	\subfigure[]{\label{fig:subfig:b}
		\includegraphics[width=0.42\linewidth]{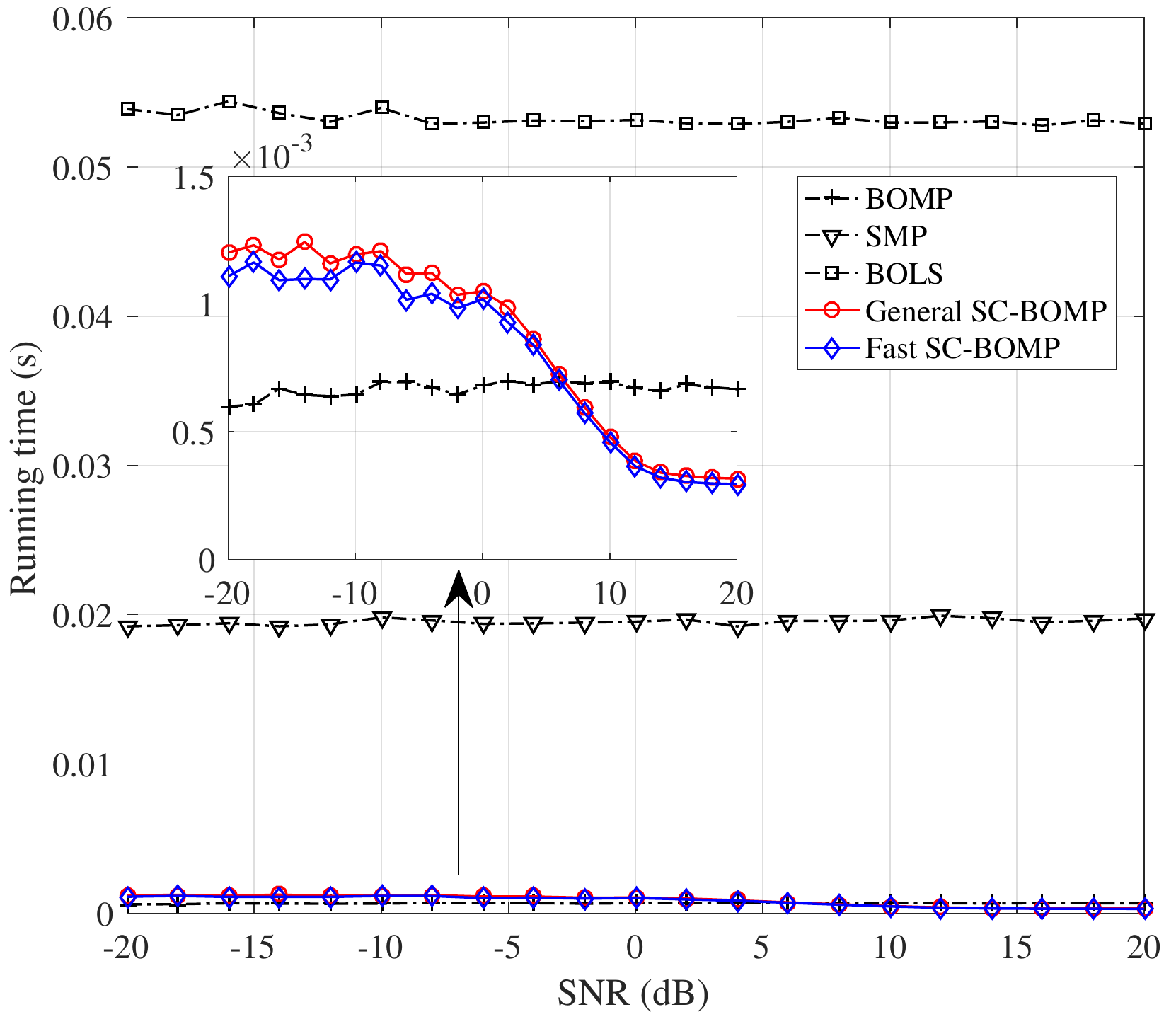}}
	\hspace{0.01\linewidth}
	\subfigure[]{\label{fig:subfig:c}
		\includegraphics[width=0.45\linewidth]{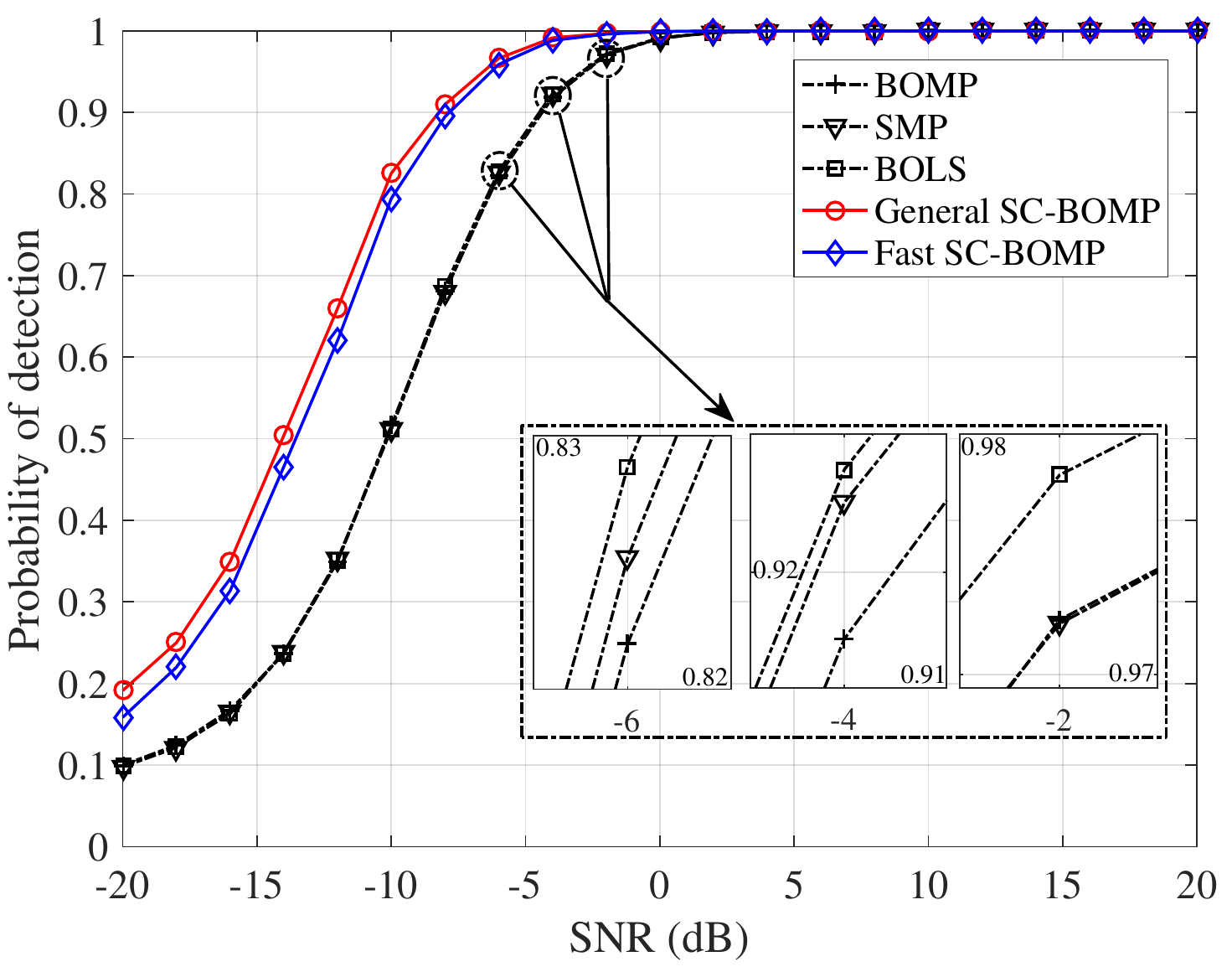}}
	\hspace{0.01\linewidth}
	\subfigure[]{\label{fig:subfig:d}
		\includegraphics[width=0.42\linewidth]{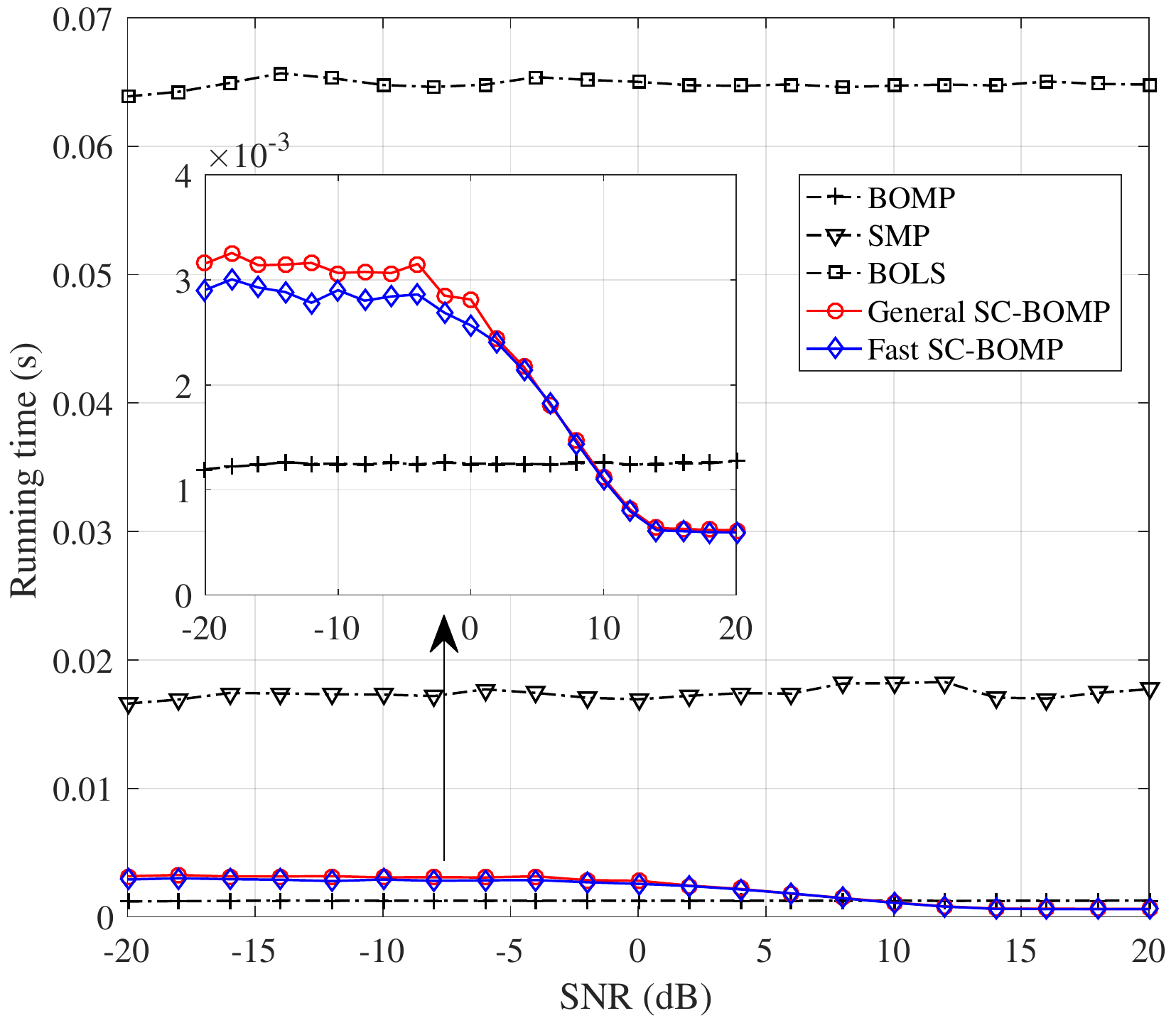}}
	\caption{Results of CSS with $k=4$, $N = 1024$. The maximum block length of (a) and (b) is $8$; the maximum block length of (c) and (d) is $16$.}
	\label{CSSsimulationres2}
\end{figure*}

In Fig. \ref{CSSsimulationres2}, the proposed SC-BOMP schemes are further compared with  subspace matching pursuit (SMP) \cite{40} and block orthogonal least squares (BOLS) \cite{8} algorithms. The compression rate of BOMP, SMP and BOLS is fixed to $1/2$. Different from that in Fig. \ref{CSSsimulationres}, the length of each non-zero block of the spectrum varies randomly within a given range. 
Unless otherwise specified, other simulation settings are the same as those in Fig. \ref{CSSsimulationres}.

It can be observed from Fig. \ref{CSSsimulationres2}, our proposed general and fast SC-BOMP schemes obtain better CSS performance than the other benchmark algorithms, while reducing the running time dynamically with the increase of SNR. Specially, in Fig. \ref{CSSsimulationres2} (a), the block length is selected within the range $[2,8]$. In low SNR environments, the general SC-BOMP performs the best followed by the fast SC-BOMP. This indicates that the proposed SC-BOMP schemes are more robust to noise effects than the other algorithms, while the performance of the latter ones is close to each other.
In high SNR environments, the CSS performance of BOMP, SMP, BOLS and the SC-BOMP schemes is all equal to 1. Moreover, the running time of both SC-BOMP schemes is shorter than the other algorithms, and the fast SC-BOMP runs faster than the general one. This reveals our proposed SC-BOMP schemes are more suitable for real-time spectrum sensing applications.

In Figs. \ref{CSSsimulationres2} (c) and (d), the block length is selected within the range $[2,16]$. With the increase of the block length range, in low SNR environments, the performance of BOMP, SMP and BOLS decreases slightly due to the increase of the overall sparsity, while our proposed SC-BOMP schemes still maintain the best performance than that of the other algorithms. This is because our proposed SC-BOMP schemes can calculate the number of measurements required for reliable recovery in real time, which guarantees the performance reliability at the expense of extra running time. Similar to Figs. \ref{CSSsimulationres2} (a) and (b), the same trend can be obtained in Figs. \ref{CSSsimulationres2} (c) and (d).

Based on the simulation results, the advantages of SC-BOMP schemes are summarized as follows: 1) As the SNR increases, SC-BOMP schemes' running time is dynamically shortened because of the reduced number of measurements; 2) the numbers of measurements in high SNR environments are just sufficiently enough for 100\% detection probabilities to reduce the waste of computing resources. 3) in low SNR environments, SC-BOMP schemes can maintain their reliability by appropriately increasing the number of measurements.

\section{Conclusion}

	This paper proposes two CSS shemes, i.e., the general SC-BOMP and the fast SC-BOMP, to achieve reliable and real-time sensing. First, the necessary number of measurements for reliable recovery is derived by developing the block MIP-based norm bounds of the matrix and noise. Second, two SCAs are proposed to adjust the computational complexity of BOMP based on the necessary number of measurements' bounds. Third, by using the two SCAs, this paper proposes two SC-BOMP schemes to moderate the wastage of sampling resources. Finally, simulation results verify that our theoretical analyses improve the existing ones in the literature, and the proposed schemes are more robust to noise effects than the benchmark schemes while adaptively reducing their complexity.

\appendices
\section{Proof of Lemma \ref{lemma4}}
\label{appendixprooflemma4}
\begin{proof}[Proof of Lemma \ref{lemma4}]
Note that $\int_{\mathbf{C}}h(x+\beta y)dx\geq \int_{\mathbf{C}}h(x+y)dx$ is equivalent to $\int_{\mathbf{C}+\beta y}h(x)dx\geq \int_{\mathbf{C}+y}h(x)dx$,
where $\mathbf{C}+\beta y$ represents a translated convex set moved by $\beta y$. The lemma can be proved that for every $u$, $V\{(\mathbf{C}+\beta y)\cap \mathbf{B}_u\}\geq V\{(\mathbf{C}+y)\cap \mathbf{B}_u\}$, where $V\{\cdot\}$ represents the volume of its target. Then, by using the similar procedures in \cite{5}, the proof is completed.
\end{proof}

\section{Proof of Lemma \ref{lemma3}}
\label{appendixlemma3}

Due to the atom selection procedure of BOMP algorithm, it is necessary to show that $\mathbf{S}^{t+1}\in\mathbf{\Omega}\backslash\mathbf{S}^t$, i.e.,
\begin{equation}\label{lemma11}
||\mathbf{D}_{\mathbf{\Omega}\backslash \mathbf{S}^t}^T\mathbf{r}^t||_{2,\infty}>||\mathbf{D}_{\bar{\mathbf{\Omega}}}^T\mathbf{r}^t||_{2,\infty}.
\end{equation}

From \cite{1}, the residual signal satisfies: $\mathbf{r}^t=\mathbf{y}-\mathbf{D}_{\mathbf{S}^t}\hat{\mathbf{x}}^t=\mathbf{P}^\bot_{\mathbf{S}^t}\mathbf{D}_{\mathbf{\Omega}\backslash \mathbf{S}^t}\mathbf{x}_{\mathbf{\Omega}\backslash \mathbf{S}^t}+\mathbf{P}^{\bot}_{\mathbf{S}^t}\mathbf{n}$,
where $\hat{\mathbf{x}}^t$ is the estimated sparse signal in the $t$-th iteration.
Then,
\begin{equation}\label{lemma13}
\begin{aligned}
&||\mathbf{D}_{\mathbf{\Omega}\backslash \mathbf{S}^t}^T\mathbf{r}^t||_{2,\infty}\\
=&||\mathbf{D}_{\mathbf{\Omega}\backslash \mathbf{S}^t}^T(\mathbf{P}^\bot_{\mathbf{S}^t}\mathbf{D}_{\mathbf{\Omega}\backslash \mathbf{S}^t}\mathbf{x}_{\mathbf{\Omega}\backslash \mathbf{S}^t}+\mathbf{P}^{\bot}_{\mathbf{S}^t}\mathbf{n})||_{2,\infty}\\
\geq&||\mathbf{D}^T_{\mathbf{\Omega}\backslash \mathbf{S}^t}\mathbf{P}^{\bot}_{\mathbf{S}^t}\mathbf{D}_{\mathbf{\Omega}\backslash \mathbf{S}^t}\mathbf{x}_{\mathbf{\Omega}\backslash \mathbf{S}^t}||_{2,\infty}-||\mathbf{D}_{\mathbf{\Omega}\backslash \mathbf{S}^t}^T\mathbf{P}_{\mathbf{S}^t}^\bot\mathbf{n}||_{2,\infty}
\end{aligned}
\end{equation}
and
\begin{equation}\label{lemma14}
\begin{aligned}
&||\mathbf{D}_{\bar{\mathbf{\Omega}}}^T\mathbf{r}^t||_{2,\infty}\\
=&||\mathbf{D}_{\bar{\mathbf{\Omega}}}^T(\mathbf{P}^\bot_{\mathbf{S}^t}\mathbf{D}_{\mathbf{\Omega}\backslash \mathbf{S}^t}\mathbf{x}_{\mathbf{\Omega}\backslash \mathbf{S}^t}+\mathbf{P}^{\bot}_{\mathbf{S}^t}\mathbf{n})||_{2,\infty}\\
\leq&||\mathbf{D}^T_{\bar{\mathbf{\Omega}}}\mathbf{P}^{\bot}_{\mathbf{S}^t}\mathbf{D}_{\mathbf{\Omega}\backslash \mathbf{S}^t}\mathbf{x}_{\mathbf{\Omega}\backslash \mathbf{S}^t}||_{2,\infty}+||\mathbf{D}_{\bar{\mathbf{\Omega}}}^T\mathbf{P}_{\mathbf{S}^t}^\bot\mathbf{n}||_{2,\infty}.
\end{aligned}
\end{equation}

Combining (\ref{lemma11}), (\ref{lemma13}) and (\ref{lemma14}), the following inequality holds:
\begin{equation}\label{lemma15}
\begin{aligned}
&||\mathbf{D}^T_{\mathbf{\Omega}\backslash \mathbf{S}^t}\mathbf{P}^{\bot}_{\mathbf{S}^t}\mathbf{D}_{\mathbf{\Omega}\backslash \mathbf{S}^t}\mathbf{x}_{\mathbf{\Omega}\backslash \mathbf{S}^t}||_{2,\infty}
-||\mathbf{D}_{\mathbf{\Omega}\backslash \mathbf{S}^t}^T\mathbf{P}_{\mathbf{S}^t}^\bot\mathbf{n}||_{2,\infty}\\
&-||\mathbf{D}_{\bar{\mathbf{\Omega}}}^T\mathbf{P}_{\mathbf{S}^t}^\bot\mathbf{n}||_{2,\infty}
>||\mathbf{D}^T_{\bar{\mathbf{\Omega}}}\mathbf{P}^{\bot}_{\mathbf{S}^t}\mathbf{D}_{\mathbf{\Omega}\backslash \mathbf{S}^t}\mathbf{x}_{\mathbf{\Omega}\backslash \mathbf{S}^t}||_{2,\infty}.
\end{aligned}
\end{equation}

The first term in the left-side of (\ref{lemma15}) can be upper bounded by the analysis in \cite{2}, i.e.,
\begin{equation}\label{lemma16}
||\mathbf{D}^T_{\mathbf{\Omega}\backslash \mathbf{S}^t}\mathbf{P}^{\bot}_{\mathbf{S}^t}\mathbf{D}_{\mathbf{\Omega}\backslash \mathbf{S}^t}\mathbf{x}_{\mathbf{\Omega}\backslash \mathbf{S}^t}||_{2,\infty}\geq\frac{||\mathbf{P}_{\mathbf{S}^t}^\bot \mathbf{D}_{\mathbf{\Omega}\backslash \mathbf{S}^t}\mathbf{x}_{\mathbf{\Omega}\backslash \mathbf{S}^t}||^2_2}{\sqrt{|\mathbf{\Omega}\backslash \mathbf{S}^t|}||\mathbf{x}_{\mathbf{\Omega}\backslash \mathbf{S}^t}||_2}.
\end{equation}
Let $\lambda'$ denote the smallest singular value of $\mathbf{P}_{\mathbf{S}^t}^\bot \mathbf{D}_{\mathbf{\Omega}\backslash \mathbf{S}^t}$. Based on Lemma 5 in \cite{3}, it is obtained that
\begin{equation}\label{lemma114}
||\mathbf{P}_{\mathbf{S}^t}^\bot \mathbf{D}_{\mathbf{\Omega}\backslash \mathbf{S}^t}\mathbf{x}_{\mathbf{\Omega}\backslash \mathbf{S}^t}||_2\geq\lambda'||\mathbf{x}_{\mathbf{\Omega}\backslash \mathbf{S}^t}||_2\geq\lambda^{\prime}_{\min}||\mathbf{x}_{\mathbf{\Omega}\backslash \mathbf{S}^t}||_2.
\end{equation}
Thus,
\begin{equation}\label{lemma115}
\begin{aligned}
&||\mathbf{D}^T_{\mathbf{\Omega}\backslash \mathbf{S}^t}\mathbf{P}^{\bot}_{\mathbf{S}^t}\mathbf{D}_{\mathbf{\Omega}\backslash \mathbf{S}^t}\mathbf{x}_{\mathbf{\Omega}\backslash \mathbf{S}^t}||_{2,\infty}\\
\geq&\frac{||\mathbf{P}_{\mathbf{S}^t}^\bot \mathbf{D}_{\mathbf{\Omega}\backslash \mathbf{S}^t}\mathbf{x}_{\mathbf{\Omega}\backslash \mathbf{S}^t}||_2\lambda^{\prime}_{\min}}{\sqrt{|\mathbf{\Omega}\backslash \mathbf{S}^t|}}\\
=&\frac{||\mathbf{P}_{\mathbf{S}^t}^\bot \mathbf{D}_{\mathbf{\Omega}\backslash \mathbf{S}^t}\mathbf{x}_{\mathbf{\Omega}\backslash \mathbf{S}^t}||_2\lambda^{\prime}_{\min}}{\sqrt{k-t}}.
\end{aligned}
\end{equation}
Now, it comes to the second term in (\ref{lemma15}). It is known that when $\nu<\frac{1}{d-1}$, there exists $i\in\mathbf{\Omega}\backslash \mathbf{S}_t$ such that
\begin{align}\label{lemma1plus1}
||\mathbf{D}^T_{\mathbf{\Omega}\backslash \mathbf{S}^t}\mathbf{P}_{\mathbf{S}^t}^{\bot}\mathbf{n}||_{2,\infty}=&||\mathbf{D}^T[i]\mathbf{P}_{\mathbf{S}^t}^{\bot}\mathbf{n}||_2
\leq||\mathbf{D}^T[i]\mathbf{P}_{\mathbf{S}^t}^{\bot}||_2||\mathbf{n}||_2\nonumber\\
\overset{(a)}{\leq}&\sqrt{1+(d-1)\nu}\times\epsilon,
\end{align}
where $(a)$ follows from Lemma \ref{lemma1}.
The upper bound of $||\mathbf{D}_{\bar{\mathbf{\Omega}}}^T\mathbf{P}_{\mathbf{S}^t}^\bot\mathbf{n}||_{2,\infty}$ in (\ref{lemma15}) is omitted which is the same as that in (\ref{lemma1plus1}). Then, by using (\ref{lemma115}) and (\ref{lemma1plus1}), the following inequality holds:
\begin{align}\label{lemma116}
&||\mathbf{D}^T_{\mathbf{\Omega}\backslash \mathbf{S}^t}\mathbf{P}^{\bot}_{\mathbf{S}^t}\mathbf{D}_{\mathbf{\Omega}\backslash \mathbf{S}^t}\mathbf{x}_{\mathbf{\Omega}\backslash \mathbf{S}^t}||_{2,\infty}-||\mathbf{D}_{\mathbf{\Omega}\backslash \mathbf{S}^t}^T\mathbf{P}_{\mathbf{S}^t}^\bot\mathbf{n}||_{2,\infty}\nonumber\\
-&||\mathbf{D}_{\bar{\mathbf{\Omega}}}^T\mathbf{P}_{\mathbf{S}^t}^\bot\mathbf{n}||_{2,\infty}-||\mathbf{D}^T_{\bar{\mathbf{\Omega}}}\mathbf{P}^{\bot}_{\mathbf{S}^t}\mathbf{D}_{\mathbf{\Omega}\backslash \mathbf{S}^t}\mathbf{x}_{\mathbf{\Omega}\backslash \mathbf{S}^t}||_{2,\infty}\nonumber\\
\geq&||\mathbf{P}_{\mathbf{S}^t}^{\bot}\mathbf{D}_{\mathbf{\Omega}\backslash \mathbf{S}^t}\mathbf{x}_{\mathbf{\Omega}\backslash \mathbf{S}^t}||_2\times\bigg(\frac{\lambda^{\prime}_{\min}}{\sqrt{k-t}}\nonumber\\
-&\frac{2\epsilon\sqrt{1+(d-1)\nu}+||\mathbf{D}^T_{\bar{\mathbf{\Omega}}}\mathbf{P}^{\bot}_{\mathbf{S}^t}\mathbf{D}_{\mathbf{\Omega}\backslash \mathbf{S}^t}\mathbf{x}_{\mathbf{\Omega}\backslash \mathbf{S}^t}||_{2,\infty}}{||\mathbf{P}_{\mathbf{S}^t}^{\bot}\mathbf{D}_{\mathbf{\Omega}\backslash \mathbf{S}^t}\mathbf{x}_{\mathbf{\Omega}\backslash \mathbf{S}^t}||_2}\bigg).
\end{align}

The following analyses derive the lower bound of the term $||\mathbf{P}^{\bot}_{\mathbf{S}^t}\mathbf{D}_{\mathbf{\Omega}\backslash \mathbf{S}^t}\mathbf{x}_{\mathbf{\Omega}\backslash \mathbf{S}^t}||_2$ in (\ref{lemma116}).
Since $||\mathbf{x}_{\mathbf{\Omega}\backslash \mathbf{S}^t}||_2\geq\sqrt{k-t}\min_{i\in\mathbf{\Omega}\setminus\mathbf{S}^t}||\mathbf{x}[i]||_2
\geq\sqrt{k-t}\min_{i\in\mathbf{\Omega}}||\mathbf{x}[i]||_2$,
\begin{equation}\label{xmaginequality2}
\begin{aligned}
||\mathbf{P}^{\bot}_{\mathbf{S}^t}\mathbf{D}_{\mathbf{\Omega}\backslash \mathbf{S}^t}\mathbf{x}_{\mathbf{\Omega}\backslash \mathbf{S}^t}||_2&\geq\lambda^{\prime}_{\min}||\mathbf{x}_{\mathbf{\Omega}\backslash \mathbf{S}^t}||_2\\
&\geq\lambda^{\prime}_{\min}\sqrt{k-t}\min_{i\in\mathbf{\Omega}}||\mathbf{x}[i]||_2.
\end{aligned}
\end{equation}

Then, let (\ref{lemma116}) be larger than zero and combine it with (\ref{xmaginequality2}), 
\begin{equation}\label{lemma117}
\begin{aligned}
||\mathbf{D}^T_{\bar{\mathbf{\Omega}}}\mathbf{g}^t||_{2,\infty}<&\frac{\lambda^{\prime}_{\min}}{\sqrt{k-t}}
-\frac{2\epsilon\sqrt{1+(d-1)\nu}}{\lambda^{\prime}_{\min}\sqrt{(k-t)}\min\limits_{i\in\mathbf{\Omega}}||\mathbf{x}[i]||_2}.
\end{aligned}
\end{equation}
The lemma then follows.
\hfill $\blacksquare$

\section{Lemma \ref{lemma6} and Proof of Theorem \ref{theoremmaindescription1}}
\label{theoremmainproof1}

In order to facilitate the following proof of Theorem \ref{theoremmaindescription1}, the subsequent Lemma \ref{lemma6} is first introduced.
Lemma \ref{lemma6} is derived based on Lemma \ref{lemma5}, which shows a lower bound on the probability of the left-side of (\ref{lemma1result}) being less than a given constant $\delta$. Without loss of generality, set $\mathbf{D}^T_{\bar{\mathbf{\Omega}}}$ as $\mathbf{D}$ in Lemma \ref{lemma6}.
\begin{lemma6}
Let $\mathbf{D}\in \mathcal{R}^{M\times N}=\mathcal{R}^{Ld\times Rd}$ be a random matrix with the atoms following i.i.d. $\mathcal{N}(0,\frac{1}{M})$ and $c$ is a positive integer. Suppose that $\mathbf{g}_i\in \mathcal{R}^{M}$ is independent with $\mathbf{D}$, satisfying $||\mathbf{g}_i||_2\leq1$ $(1\leq i\leq c)$. Then, for $\delta_i\geq\frac{d(1+\gamma)}{M}$ $(\gamma>0)$, the following inequality holds:
\begin{equation}\label{Lemma5main}
\begin{aligned}
&{\rm P}\bigg(\bigcap_{i=1}^c(||\mathbf{D}^T\mathbf{g}_i||_{2,\infty}\leq\sqrt{\delta_i})\bigg)\\
\geq&\prod^c_{i=1}\Bigg(1-\frac{(\frac{M\delta_i}{d})^{\frac{d}{2}}}{\sqrt{\pi d}(\frac{M\delta_i}{d}-1)}e^{\frac{d}{2}(1-\frac{M\delta_i}{d})}\Bigg)^R.
\end{aligned}
\end{equation}
\label{lemma6}
\end{lemma6}

\begin{proof}
See Appendix \ref{proofofthelemmas}.
\end{proof}

\begin{proof}[Proof of Theorem \ref{theoremmaindescription1}]
To prove the theorem, it is necessary to show that $\mathbf{S}^t\subseteq\mathbf{\Omega}$ and $|\mathbf{S}^t|=k$ $(0\leq t\leq k-1)$. Therefore, by Lemma \ref{lemma3} and induction, it is equivalent to show that (\ref{lemma1result}) holds for $0\leq t\leq k-1$.
For simplicity, denote the event $\mathbf{E}(t,\lambda^{\prime}_{\min})=\{||\mathbf{D}^T_{\bar{\mathbf{\Omega}}}\mathbf{g}^t||_{2,\infty}<\eta(k-t,\lambda^{\prime}_{\min})\},\; 0\leq t\leq k-1$,
where $\eta(\mathcal{P},\mathcal{Q})=\frac{\mathcal{Q}}{\sqrt{\mathcal{P}}}
-\frac{2\epsilon\sqrt{1+(d-1)\nu}}{\mathcal{Q}\sqrt{\mathcal{P}}\min\limits_{i\in\mathbf{\Omega}}||\mathbf{x}[i]||_2}$.
For any
\begin{equation}\label{theorem1plus1}
\begin{aligned}
&0<\zeta<1-\sqrt{\frac{kd}{M}}
-\sqrt{\frac{d(c_0+1)k}{4M}}\\
&-\sqrt{\frac{d(c_0+1)k}{4M}
+\frac{2\epsilon\sqrt{1+(d-1)\nu}}{\min\limits_{i\in\mathbf{\Omega}}||\mathbf{x}[i]||_2}},
\end{aligned}
\end{equation}
based on (\ref{eigvaluedefinition}),
\begin{equation}\label{eigvaluevalue}
\begin{aligned}
&\tilde{\lambda}=1-\sqrt{\frac{kd}{M}}-\zeta
>\sqrt{\frac{d(c_0+1)k}{4M}}\\
&+\sqrt{\frac{d(c_0+1)k}{4M}+\frac{2\epsilon\sqrt{1+(d-1)\nu}}{\min\limits_{i\in\mathbf{\Omega}}||\mathbf{x}[i]||_2}}.
\end{aligned}
\end{equation}

Then, due to (\ref{yetadefinition}) and (\ref{eigvaluevalue}), 
\begin{equation}\label{eigvaluevalu2}
\begin{aligned}
\eta(t,\tilde{\lambda})=\frac{\tilde{\lambda}}{\sqrt{t}}
-\frac{2\epsilon\sqrt{1+(d-1)\nu}}{\tilde{\lambda}\sqrt{t}\min\limits_{i\in\mathbf{\Omega}}||\mathbf{x}[i]||_2}>\sqrt{\frac{d(c_0+1)}{M}}.
\end{aligned}
\end{equation}

Therefore, 
\begin{equation}\label{eigvaluevalu3}
\frac{M\eta^2(t,\tilde{\lambda})}{d}-1>c_0.
\end{equation}

For clarity, denote function $g(x,d)$ as $g(x,d)=\frac{(x+1)^{\frac{d}{2}}}{\sqrt{\pi d}x}e^{-\frac{d}{2}x}$,
where $x>0$ and $d$ is positive integer.
After some basic calculations, it is obtained that $\left\{\begin{matrix}\frac{\partial g(x,d)}{\partial x}<0,\\
	\frac{\partial g(x,d)}{\partial d}<0.
\end{matrix}\right.$
This indicates that the maximum value of the function $g(x,d)$ is taken at the minimum available values of the arguments.
Let $x(t,\tilde{\lambda})=\frac{M\eta^2(t,\tilde{\lambda})}{d}-1$ and due to (\ref{positivesolution}) and (\ref{eigvaluevalu3}), $g(x(t,\tilde{\lambda}),d)<1,\;1\leq t\leq k$.
Based on Lemma \ref{lemma3}, Lemma \ref{lemma6} and the fact that ${\rm P}\Big(\bigcap_{t=0}^{k-1}(||\mathbf{D}^T_{\bar{\mathbf{\Omega}}}\mathbf{g}^t||_{2,\infty}<\eta(t))\Big)
={\rm P}\Big(\bigcap_{t=0}^{k-1}(||\mathbf{D}^T_{\bar{\mathbf{\Omega}}}\mathbf{g}^t||_{2,\infty}\leq\eta(t))\Big)$,
\begin{align}
&{\rm P}(\mathbf{E})\nonumber
\geq{\rm P}\Bigg(\bigcap_{t=0}^{k-1}\mathbf{E}(k-t,\lambda^{\prime}_{\min})\Bigg)\nonumber\\
\geq&{\rm P}\Bigg(\bigcap_{t=0}^{k-1}\mathbf{E}(k-t,\lambda^{\prime}_{\min}),\lambda^{\prime}_{\min}\geq\tilde{\lambda}\Bigg)\nonumber\\
=&{\rm P}\Bigg(\bigcap_{t=0}^{k-1}\mathbf{E}(k-t,\lambda^{\prime}_{\min})|\lambda^{\prime}_{\min}\geq\tilde{\lambda}\Bigg)\times{\rm P}\big(\lambda^{\prime}_{\min}\geq\tilde{\lambda}\big)\nonumber\\
=&{\rm P}\Bigg(\bigcap_{t=0}^{k-1}\Big(||\mathbf{D}^T_{\bar{\mathbf{\Omega}}}\mathbf{g}_t||_{2,\infty}<\eta(k-t,\tilde{\lambda})\Big)\Bigg)\nonumber
\times{\rm P}(\lambda^{\prime}_{\min}\geq\tilde{\lambda})\nonumber\\
\geq&\prod_{t=0}^{k-1}\Bigg(1-\frac{(\frac{M\eta^2(k-t,\tilde{\lambda})}{d})^{\frac{d}{2}}}{\sqrt{\pi d}(\frac{M\eta^2(k-t,\tilde{\lambda})}{d}-1)}e^{\frac{d}{2}(1-\frac{M\eta^2(k-t,\tilde{\lambda})}{d})}\Bigg)^{R-k}\nonumber\\
&\times\Big(1-e^{-\frac{\zeta^2M}{2}}\Big)\nonumber\\
=&\prod_{t=0}^{k-1}\Bigg(1-\frac{(\frac{M\eta^2(t,\tilde{\lambda})}{d})^{\frac{d}{2}}}{\sqrt{\pi d}(\frac{M\eta^2(t,\tilde{\lambda})}{d}-1)}e^{\frac{d}{2}(1-\frac{M\eta^2(t,\tilde{\lambda})}{d})}\Bigg)^{R-k}\nonumber\\
&\times\Big(1-e^{-\frac{\zeta^2M}{2}}\Big).\label{theorem12}
\end{align}

Finally, the proof of Theorem \ref{theoremmaindescription1} is completed.
\end{proof}

\section{Lemma \ref{lemma7} and Proof of Theorem \ref{theoremmaindescription2}}
\label{theoremmainproof2}

\begin{lemma7}
Let $h(x)=\frac{(x+1)^{\frac{d}{2}}}{x}e^{-\frac{d}{2}x}$ $(d>2)$. Suppose that $2\leq x_1\leq x_2\leq\cdots\leq x_c$ for an integer $c$, then $\sum_{i=1}^{c}h(x_i)\leq\frac{(x_c+1)^{\frac{d}{2}}\sum_{i=1}^ce^{-\frac{d}{2}x_i}}{x_c}$.
\label{lemma7}
\end{lemma7}
\begin{proof}
See Appendix \ref{proofofthelemmas}.
\end{proof}

Then, the following analyses exploit Theorem \ref{theoremmaindescription1} and Lemma~\ref{lemma7} to prove Theorem \ref{theoremmaindescription2}.

\begin{proof}[Proof of Theorem \ref{theoremmaindescription2}]
Let
\begin{equation}\label{proofoftheorem20}
\zeta_0=\sqrt{\frac{2\ln(N/\omega)}{M}},\:\tilde{\lambda}_0=1-\sqrt{\frac{kd}{M}}-\zeta_0,
\end{equation}
\begin{equation}\label{proofoftheorem21}
\begin{aligned}
&{\rm P}(\mathbf{E}_{\zeta_0})=\Big(1-e^{-\frac{\zeta_0^2M}{2}}\Big)\\
&\times\prod_{t=1}^{k}\Bigg(1-\frac{(\frac{M\eta^2(t,\tilde{\lambda}_0)}{d})^{\frac{d}{2}}}{\sqrt{\pi d}(\frac{M\eta^2(t,\tilde{\lambda}_0)}{d}-1)}e^{\frac{d}{2}(1-\frac{M\eta^2(t,\tilde{\lambda}_0)}{d})}\Bigg)^{R-k}.
\end{aligned}
\end{equation}
The following proofs first prove that
\begin{equation}\label{proofoftheorem2firstshow}
\begin{aligned}
\zeta_0<&1-\sqrt{\frac{kd}{M}}
-\sqrt{\frac{d(c_0+1)k}{4M}}\\
&-\sqrt{\frac{d(c_0+1)k}{4M}
+\frac{2\epsilon\sqrt{1+(d-1)\nu}}{\min\limits_{i\in\mathbf{\Omega}}||\mathbf{x}[i]||_2}},
\end{aligned}
\end{equation}
and then demonstrate that ${\rm P}(\mathbf{E}_{\zeta_0})\geq1-\frac{\omega}{N}-\frac{\omega}{\sqrt{\pi d}}$.

First, based on (\ref{proofoftheorem20}) and (\ref{proofoftheorem2firstshow}), 
\begin{equation}\label{proofoftheorem2theinequality}
\begin{aligned}
&\zeta_0+\sqrt{\frac{kd}{M}}+\sqrt{\frac{d(c_0+1)k}{4M}}\\
&+\sqrt{\frac{d(c_0+1)k}{4M}
+\frac{2\epsilon\sqrt{1+(d-1)\nu}}{\min\limits_{i\in\mathbf{\Omega}}||\mathbf{x}[i]||_2}}\\
\leq&\frac{1}{\sqrt{M}}\Bigg(\sqrt{2\ln\Big(\frac{N}{\omega}\Big)}+\sqrt{kd}+\sqrt{\frac{d(c_0+1)k}{4}}\\
&+\sqrt{\frac{d(c_0+1)k}{4}+\frac{2\epsilon M\sqrt{1+(d-1)\nu}}{\min\limits_{i\in\mathbf{\Omega}}||\mathbf{x}[i]||_2}}\Bigg)\\
=&\frac{1}{\sqrt{M}}\big(\alpha_1+\sqrt{\alpha_2+\alpha_3M}\big),
\end{aligned}
\end{equation}

Observe that if
\begin{equation}\label{proofoftheorem2theinequality2}
\begin{aligned}
M\geq&\bigg(\sqrt{\frac{2}{kd}}+\sqrt{\frac{1}{\ln(\frac{N}{\omega})}}+\sqrt{\frac{\alpha_2}{kd\ln(\frac{N}{\omega})}}\\
&-\sqrt{\frac{\alpha_2+\alpha_1^2\alpha_3-\alpha_2\alpha_3}{kd\ln(\frac{N}{\omega})}}\bigg)^2kd\ln\Big(\frac{N}{\omega}\Big),
\end{aligned}
\end{equation}
(\ref{proofoftheorem2theinequality}) is less than or equal to 1 and thus (\ref{proofoftheorem2firstshow}) holds.

Secondly,
it is known from \cite{14} that
\begin{equation}\label{theorem21}
\prod_{t=1}^{k}(1-a_t\rho)\geq1-\bigg(\sum_{t=1}^{k}a_t\bigg)\rho,\; a_t\geq0, \;\rho\geq0.
\end{equation}

In order to simplify the derivation process, denote $\frac{M\eta^2(t,\tilde{\lambda}_0)}{d}-1=q(t)$ $(1\leq t\leq k)$. Due to the bound of (\ref{eigvaluevalu3}), $q(t)$ satisfies the domain definition of Lemma \ref{lemma7} with respect to $x$. Hence, based on Theorem \ref{theoremmaindescription1}, Lemma \ref{lemma7} and (\ref{theorem21}), the following inequality holds:
\begin{equation}\label{theorem22}
\begin{aligned}
{\rm P}(\mathbf{E}_{\zeta_0})
\geq&1-e^{-\frac{\zeta_0^2M}{2}}-\frac{R-k}{\sqrt{\pi d}}\sum_{t=1}^k\frac{(q(t)+1)^{\frac{d}{2}}}{q(t)}e^{-\frac{d}{2}q(t)}\\
\geq&1-e^{-\frac{\zeta_0^2M}{2}}-\frac{R-k}{\sqrt{\pi d}}\times\frac{(q(k)+1)^{\frac{d}{2}}}{q(k)}\sum_{t=1}^ke^{-\frac{d}{2}q(t)}.
\end{aligned}
\end{equation}

Then, it is necessary to obtain the upper bound of the last two terms of (\ref{theorem22}).
Note that $e^{-\frac{\zeta_0^2M}{2}}=e^{-\ln(\frac{N}{\omega})}=\frac{\omega}{N}$.
Now it remains to derive the upper bound of the last term.
Since $\frac{\partial q(t)}{\partial t}<0$, based on (\ref{theorem222main1}), when $\xi>1$, 
\begin{align}\label{xibounds}
\xi&=\frac{R-k}{\ln(N)}\times4^{2d}ke^{-\frac{d}{2}}
\overset{(a)}{\geq}\frac{R-k}{\ln(N)}(q(k)+1)^{2d}ke^{-\frac{d}{2}q(k)}\nonumber\\
&\overset{(b)}{\geq}\frac{R-k}{\ln(N)}(q(k)+1)^{2d}\sum_{t=1}^{k}e^{-\frac{d}{2}q(t)},
\end{align}
where $(a)$ holds because the maximum value of the function $(q(k)+1)^de^{-\frac{d}{2}q(k)}$ is obtained at $q(k)=3$ and $(b)$ is because $q(t)$ $(1\leq t\leq k)$ is a monotonically decreasing function.
Therefore, if
\begin{equation}\label{proofoftheorem2theinequality2}
q(k)\geq\sqrt[\frac{3d}{2}+1]{ \frac{\xi\ln(N)}{\omega}},
\end{equation}
the following inequality holds:
\begin{equation}\label{proofoftheorem2theinequality5}
\begin{aligned}
&\frac{R-k}{\sqrt{\pi d}}\times\frac{(q(k)+1)^{\frac{d}{2}}}{q(k)}\sum_{t=1}^ke^{-\frac{d}{2}q(t)}\\
\leq&\frac{R-k}{\sqrt{\pi d}}\times\frac{(q(k)+1)^{2d}}{q(k)^{\frac{3d}{2}+1}}\sum_{t=1}^ke^{-\frac{d}{2}q(t)}\\
\leq&\frac{(R-k)\omega}{\sqrt{\pi d}\xi \ln(N)}\times(q(k)+1)^{2d}\sum_{t=1}^ke^{-\frac{d}{2}q(t)}
\leq\frac{\omega}{\sqrt{\pi d}}.
\end{aligned}
\end{equation}
The last inequality follows from (\ref{theorem222main1}) and (\ref{xibounds}).
Therefore, the prove of the upper bound of the last term in (\ref{theorem22}) changes into the prove of (\ref{proofoftheorem2theinequality2}). By some calculations, the following inequality is obtained:
\begin{equation}\label{theorem23}
\begin{aligned}
(\min\limits_{i\in\mathbf{\Omega}}||\mathbf{x}[i]||_2&-\beta_1)u^2+(-2\beta_1\beta_2-\beta_3)u-\beta_1\beta_2^2\geq0,
\end{aligned}
\end{equation}
where $u=\sqrt{M}-\sqrt{kd}-\sqrt{2\ln(\frac{N}{\omega})}$.
The positive solution of (\ref{theorem23}) with respect to $M$ is consistent with the second term in (\ref{theoremmain2noisecase}).
Based on the above analyses, the proof is completed.
\end{proof}

\section{Proof of Lemma \ref{lemma6} and Lemma \ref{lemma7}}
\label{proofofthelemmas}

\begin{proof}[Proof of Lemma \ref{lemma6}]
Since $\mathbf{g}_i$ $(1\leq i\leq c)$ is independent of $\mathbf{D}$,  denote $\mathbf{D}_j^T\mathbf{g}_i\sim \mathcal{N}(0,\sigma^2)$ $(j\in\bar{\mathbf{\Omega}})$, where $\sigma^2=\frac{||\mathbf{g}_i||_2^2}{M}\leq\frac{1}{M}$. It is known that $X=\frac{||\mathbf{D}^T\mathbf{g}_i||^2_{2,\infty}}{\sigma^2}$ obeys chi-square distribution with degree of freedom $d$, i.e., $\chi^2_d$, which satisfies the assumptions of Lemma \ref{lemma5}. Then, 
\begin{align}
&{\rm P}\bigg(\bigcap_{i=1}^c(||\mathbf{D}^T\mathbf{g}_i||_{2,\infty}\leq\sqrt{\delta_i})\bigg)\nonumber\\
=&{\rm P}(0\leq||\mathbf{D}^T\mathbf{g}_1||_{2,\infty}\leq\sqrt{\delta_1},\cdots,0\leq||\mathbf{D}^T\mathbf{g}_c||_{2,\infty}\leq\sqrt{\delta_c})\nonumber\\
=&{\rm P}(0\leq||\mathbf{D}^T[1]\mathbf{g}_1||^2_2\leq\delta_1,\cdots,0\leq||\mathbf{D}^T[R]\mathbf{g}_1||^2_2\leq\delta_1,\cdots,\nonumber\\
&0\leq||\mathbf{D}^T[1]\mathbf{g}_c||^2_2\leq\delta_c,\cdots,0\leq||\mathbf{D}^T[R]\mathbf{g}_c||^2_2\leq\delta_c)\nonumber\\
\geq&{\rm P}(0\leq||\mathbf{D}^T[1]\mathbf{g}_1||^2_2\leq\delta_1)\nonumber
\times\cdots\times{\rm P}(0\leq||\mathbf{D}^T[R]\mathbf{g}_1||^2_2\leq\delta_1)\nonumber\\
&\times\cdots\times{\rm P}(0\leq||\mathbf{D}^T[1]\mathbf{g}_c||^2_2\leq\delta_c)\nonumber\\
&\times\cdots\times{\rm P}(0\leq||\mathbf{D}^T[R]\mathbf{g}_c||^2_2\leq\delta_c)\nonumber\\
=&\prod_{i=1}^{c}\prod_{j=1}^{R}{\rm P}\bigg(0\leq\frac{||\mathbf{D}^T[j]\mathbf{g}_i||_2^2}{\sigma^2}\leq\frac{\delta_i}{\sigma^2}\bigg).\label{lemma51}
\end{align}

It remains to derive the lower bound of the probability ${\rm P}(0\leq\frac{||\mathbf{D}^T[j]\mathbf{g}_i||_2^2}{\sigma^2}\leq\frac{\delta_i}{\sigma^2})$, for $1\leq i\leq c$, $1\leq j\leq R$. Based on Lemma 4 in \cite{7}, 
\begin{align}\label{lemma52}
&{\rm P}\bigg(0\leq\frac{||\mathbf{D}^T[j]\mathbf{g}_i||_2^2}{\sigma^2}\leq\frac{\delta_i}{\sigma^2}\bigg)
=1-{\rm P}\bigg(\frac{||\mathbf{D}^T[j]\mathbf{g}_i||_2^2}{\sigma^2}>\frac{\delta_i}{\sigma^2}\bigg)\nonumber\\
&\geq1-\frac{(\frac{\delta_i}{\sigma^2d})^{\frac{d}{2}}}{\sqrt{\pi d}(\frac{\delta_i}{\sigma^2d}-1)}e^{\frac{d}{2}(1-\frac{\delta_i}{\sigma^2d})},
\end{align}
where $\delta_i>(1+\gamma)d\sigma^2$ $(\gamma>0)$ due to the restrictions on the use of Lemma 4 in \cite{7}. Since $\sigma^2\leq\frac{1}{M}$, then $\delta_i>\frac{(1+\gamma)d}{M}$.
For $||\mathbf{g}_i||_2=1$, $\sigma^2=\frac{1}{M}$. Then (\ref{lemma52}) changes into:
\begin{equation}\label{lemma53}
\begin{aligned}
&{\rm P}\bigg(0\leq\frac{||\mathbf{D}^T[j]\mathbf{g}_i||_2^2}{\sigma^2}\leq\frac{\delta_i}{\sigma^2}\bigg)\\
\geq&
1-\frac{(\frac{M\delta_i}{d})^{\frac{d}{2}}}{\sqrt{\pi d}(\frac{M\delta_i}{d}-1)}e^{\frac{d}{2}(1-\frac{M\delta_i}{d})}.
\end{aligned}
\end{equation}

Now, it comes to the case that $||\mathbf{g}_i||_2<1$. By normalizing $\mathbf{g}_i$, 
\begin{equation}\label{lemma54}
\begin{aligned}
&{\rm P}\bigg(0\leq\frac{||\mathbf{D}^T[j]\mathbf{g}_i||_2^2}{\sigma^2}\leq\frac{\delta_i}{\sigma^2}\bigg)\\
=&{\rm P}\bigg(0\leq\frac{||\mathbf{D}^T[j]\mathbf{g}_i||_2^2}{\tilde{\sigma}^2||\mathbf{g}_i||_2^2}\leq\frac{\delta_i}{\tilde{\sigma}^2||\mathbf{g}_i||_2^2}\bigg)\\
\geq&{\rm P}\bigg(0\leq\frac{||\mathbf{D}^T[j]\mathbf{g}_i||_2^2}{\tilde{\sigma}^2||\mathbf{g}_i||_2^2}\leq\frac{\delta_i}{\tilde{\sigma}^2}\bigg)\\
\geq&1-\frac{(\frac{M\delta_i}{d})^{\frac{d}{2}}}{\sqrt{\pi d}(\frac{M\delta_i}{d}-1)}e^{\frac{d}{2}(1-\frac{M\delta_i}{d})},
\end{aligned}
\end{equation}
where $\tilde{\sigma}^2$ is the variance of the variable $\frac{||\mathbf{D}^T[j]\mathbf{g}_i||_2^2}{||\mathbf{g}_i||_2^2}$.
This completes the proof of Lemma \ref{lemma6}.
\end{proof}

\begin{proof}[Proof of Lemma \ref{lemma7}]
Let $\bar{h}(x)=\frac{(x+1)^{\frac{d}{2}}}{x}$.
Since the derivative of $\bar{h}(x)$ with respect to $x$ $(x\geq2)$ satisfies: $\bar{h}^{\prime}(x)=\frac{\frac{d}{2}(x+1)^{\frac{d}{2}-1}x-(x+1)^{\frac{d}{2}}}{x^2}\geq0 \quad(d>2)$,
\begin{equation}\label{lemma73}
\begin{aligned}
&\sum_{i=1}^{c}h(x_i)=\sum_{i=1}^{c}(e^{-\frac{d}{2}x_i}\bar{h}(x_i))\leq\sum_{i=1}^{c}(e^{-\frac{d}{2}x_i}\bar{h}(x_c))\\
=&\frac{(x_c+1)^{\frac{d}{2}}\sum_{i=1}^ce^{-\frac{d}{2}x_i}}{x_c}.
\end{aligned}
\end{equation}

The proof is thus completed.
\end{proof}

\ifCLASSOPTIONcaptionsoff
  \newpage
\fi

\bibliographystyle{ieeetran}
\bibliography{SPECTRUM_SENSING}

\begin{IEEEbiography}[{\includegraphics[width=1in,height=1.25in,clip,keepaspectratio]{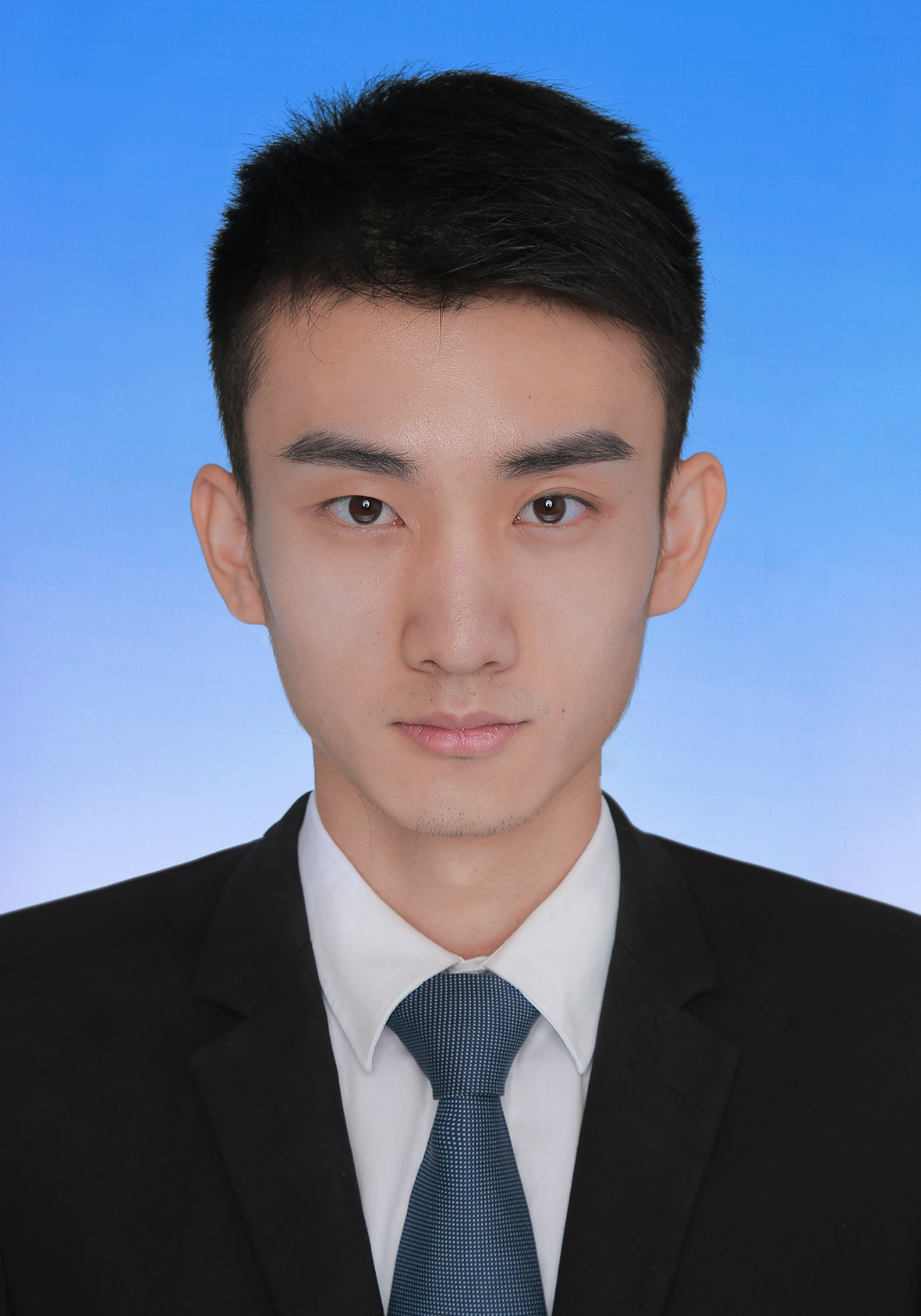}}]{Liyang Lu} (Graduate Student Member, IEEE)
	received the B.S. degree in communication engineering from Beijing University of Posts and Telecommunications (BUPT), China, in 2017. He is currently pursuing the Ph.D. degree in information and communication engineering at BUPT. His area of research includes compressive sensing, cognitive radios, integrated sensing and communication, sparse representation-based classification and signal optimization.
\end{IEEEbiography}

\begin{IEEEbiography}[{\includegraphics[width=1in,height=1.25in,clip,keepaspectratio]{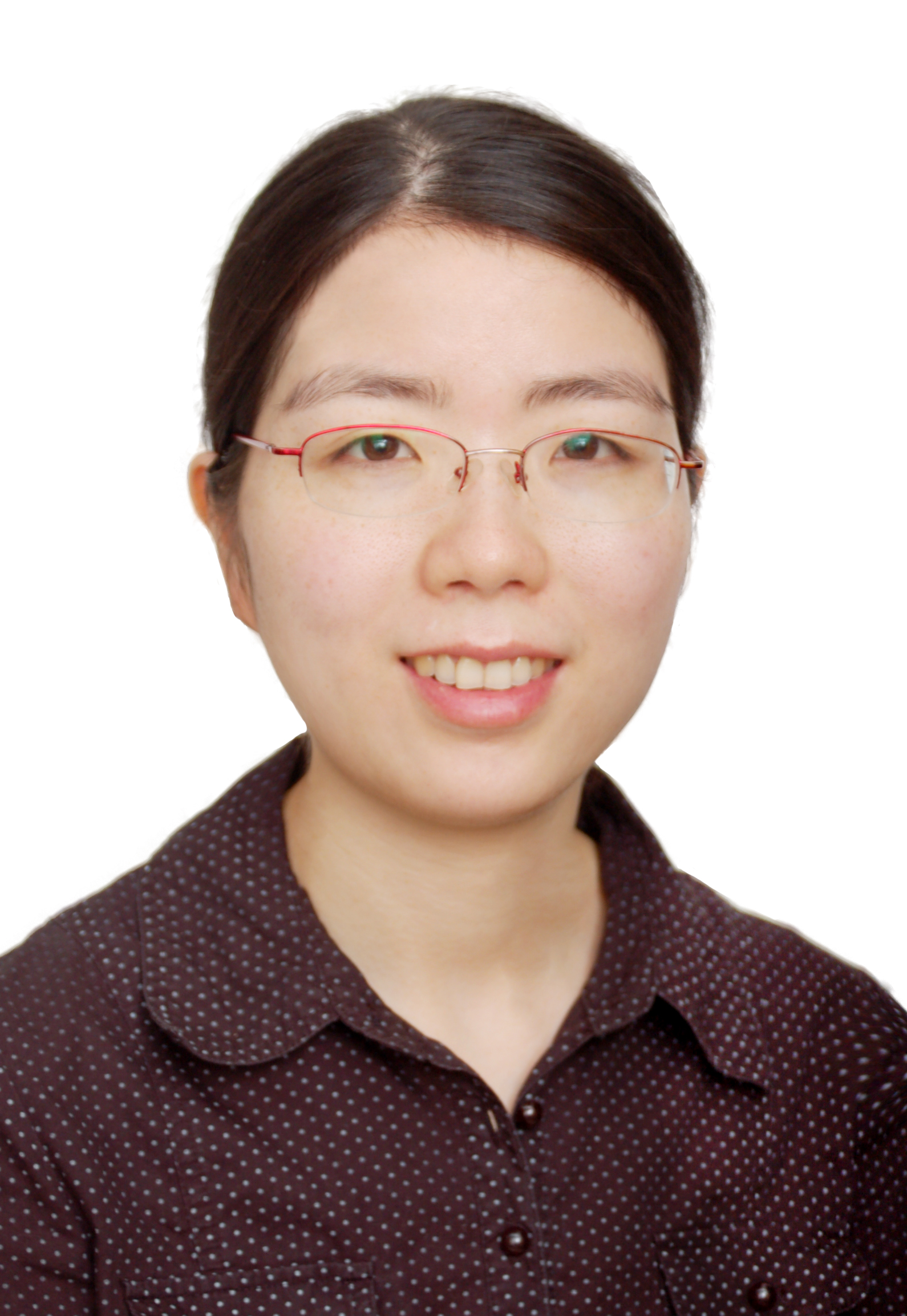}}]{Wenbo Xu} (Member, IEEE)
	received the B.S. degree from School of Information Engineering in Beijing University of Posts and Telecommunications (BUPT), China, in 2005, and the Ph.D. degree from School of Information and Communication Engineering in BUPT, China, 2010. Since 2010, she has been with BUPT, where she is currently the professor in the School of Artificial Intelligence. Her current research interests include sparse signal processing, machine learning and signal processing in wireless networks.
\end{IEEEbiography}

\begin{IEEEbiography}[{\includegraphics[width=1in,height=1.25in,clip,keepaspectratio]{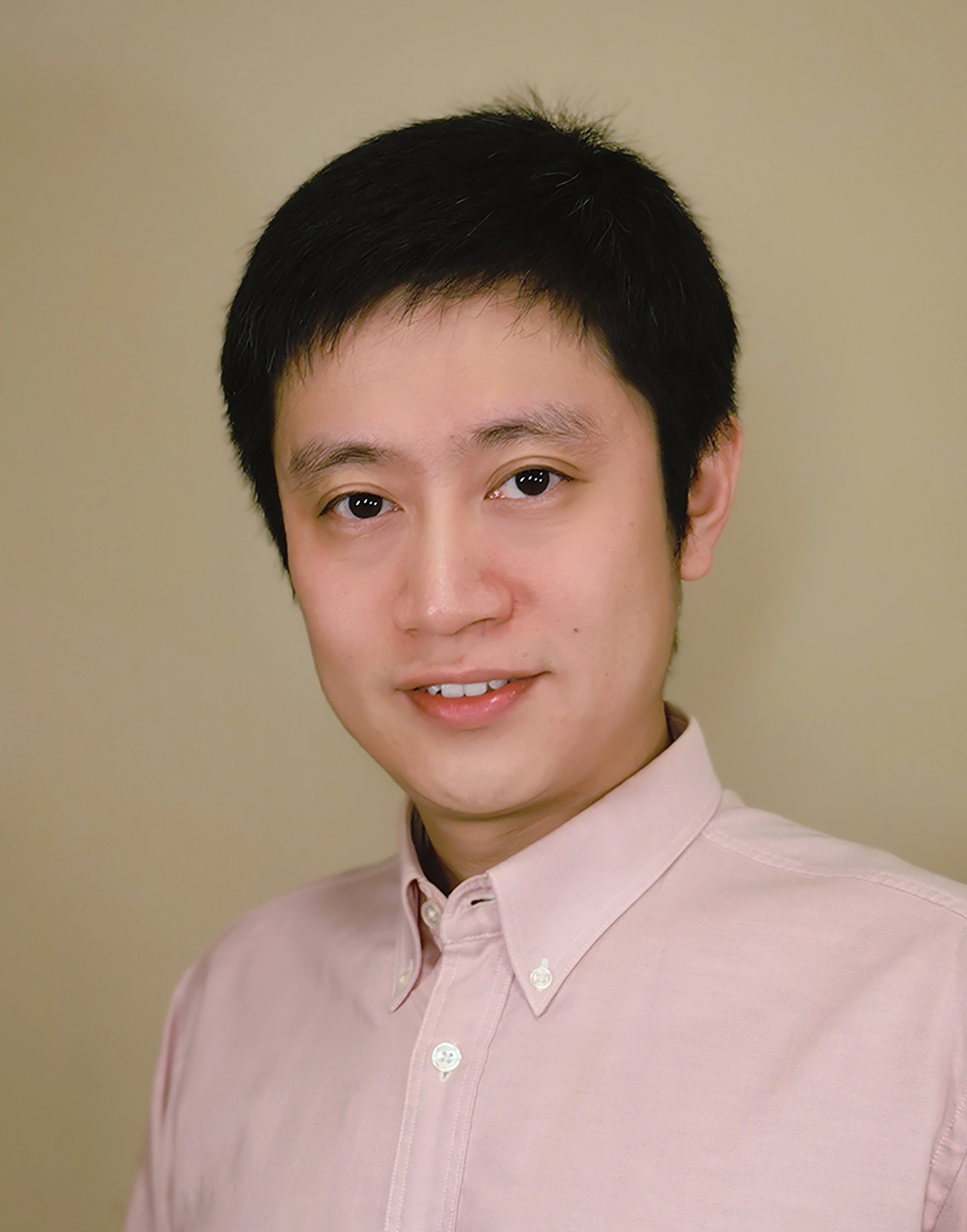}}]{Yue Wang} (Senior Member, IEEE)
	received the Ph.D. degree in communication and information system from the School of Information and Communication Engineering, Beijing University of Posts and Telecommunications, Beijing, China, in 2011. He is currently a Research Assistant Professor with Electrical and Computer Engineering Department, George Mason University, Fairfax, VA, USA, where he was a Postdoctoral Researcher. Prior to that, he was a Senior Engineer with Huawei Technologies Co., Ltd., China. From 2009 to 2011, he was a Visiting Ph.D. Student with Electrical and Computer Engineering Department, Michigan Technological University, Houghton, MI, USA. His general interests include signal processing, wireless communications, machine learning, and their applications in cyber physical systems. His specific research focuses on compressive sensing, massive MIMO, millimeter-wave communications, cognitive radios, DoA estimation, high-dimensional data analysis, and distributed optimization and learning.
\end{IEEEbiography}

\begin{IEEEbiography}[{\includegraphics[width=1in,height=1.25in,clip,keepaspectratio]{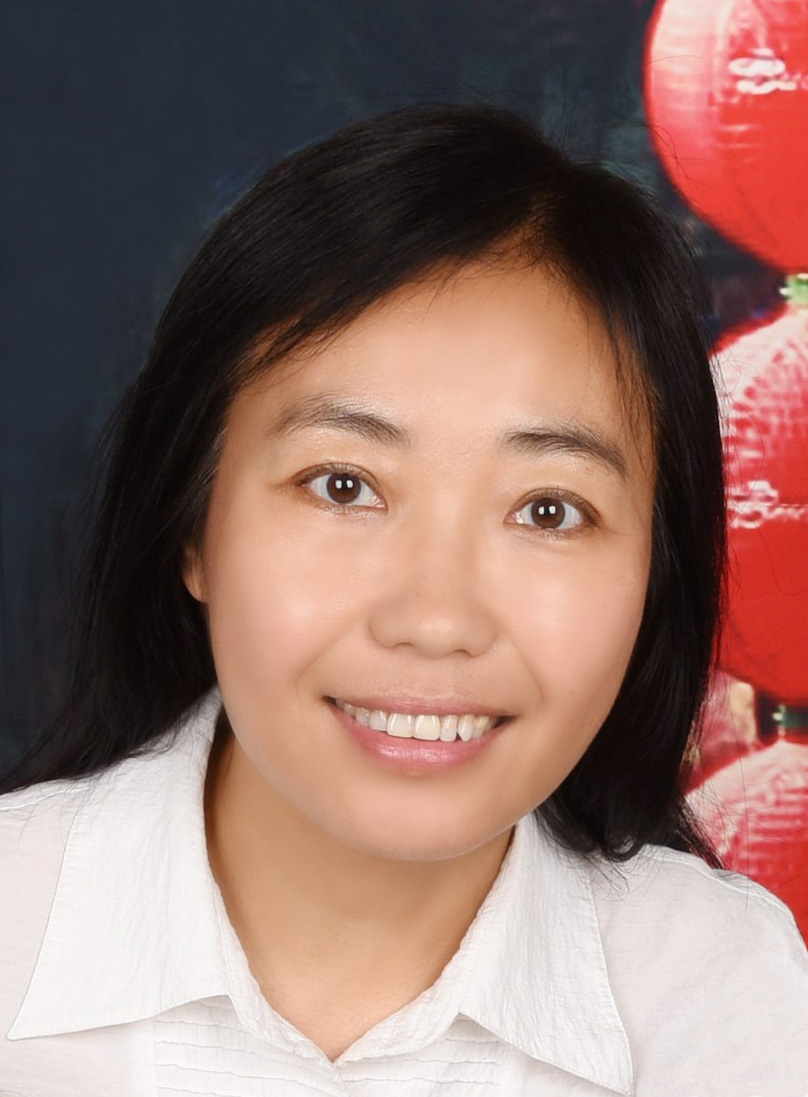}}]{Zhi Tian} (Fellow, IEEE)
	is currently a Professor with the Electrical and Computer Engineering Department of George Mason University, Fairfax, VA, USA, since 2015. Prior to that, she was on the Faculty of Michigan Technological University, Houghton, MI, USA, from 2000 to 2014. She served as a Program Director at the U.S. National Science Foundation from 2012 to 2014. Her research interest lies in the areas of statistical signal processing, wireless communications, and estimation and detection theory. Her current research focuses on compressed sensing for random processes, statistical inference of network data, distributed network optimization and learning, and millimeter-wave communications. She was an IEEE Distinguished Lecturer for both the IEEE Communications Society and the IEEE Vehicular Technology Society. She served as Associate Editor for IEEE TRANSACTIONS ON WIRELESS COMMUNICATIONS and IEEE TRANSACTIONS ON SIGNAL PROCESSING. She received the IEEE Communications Society TCCN Publication Award in 2018. She was the Chair of the IEEE Signal Processing Society Big Data Special Interest Group and the General Co-Chair of the 2016 IEEE GlobalSIP Conference, and is the Unclassified Technical Program Co-Chair of the 2022 IEEE MILCOM Conference. She was a Member-at-Large of the Board of Governors of the IEEE Signal Processing Society (2019-2021). 
\end{IEEEbiography}

\end{document}